\documentclass[preprint*]{JHEP3} 

\JHEPspecialurl{http://jhep.sissa.it/JOURNAL/JHEP3.tar.gz}

\usepackage{epsfig,multicol}
\usepackage{amsmath,amssymb,bm}

\newcommand\fverb{\setbox\pippobox=\hbox\bgroup\verb}
\newcommand\fverbdo{\egroup\medskip\noindent%
                        \fbox{\unhbox\pippobox}\ }
\newcommand\fverbit{\egroup\item[\fbox{\unhbox\pippobox}]}
\newbox\pippobox

\title{
A construction of the Glashow-Weinberg-Salam model 
on the lattice 
with exact gauge invariance
}

\author{Daisuke Kadoh \\ Center for Computational Sciences, University of Tsukuba, 
Ibaraki  305-8571, Japan\\
E-mail: \email{kadoh@ccs.tsukuba.ac.jp}}
\author{Yoshio Kikukawa \\ Institute of Physics, University of Tokyo, Tokyo 153-8902, Japan \\
E-mail: \email{kikukawa@hep1.c.u-tokyo.ac.jp}}

\preprint{
                 UT-KOMABA/07-17\\September 2007}      

\abstract{
We present a gauge-invariant and non-perturbative construction of the 
Glashow-Weinberg-Salam model on the lattice, based on the lattice Dirac operator satisfying 
the Ginsparg-Wilson relation.
Our construction covers all SU(2) topological sectors with vanishing 
U(1) magnetic flux and would be usable for a description of the baryon number non-conservation. 
In infinite volume, it provides  a gauge-invariant regularization of the 
electroweak theory 
to all orders of  perturbation theory. 
First we formulate the reconstruction theorem which asserts that
if there exists a set of local currents satisfying cetain properties, 
it is possible to reconstruct  the fermion measure which depends smoothly 
on the gauge fields  and fulfills the fundamental requirements such as locality, 
gauge-invariance and lattice symmetries.
Then we give a closed formula of the local currents required for the reconstruction theorem.
}


\begin{document} 


%
\section{Introduction}
\label{sec:intro} 
There are several interesting possibilities in the dynamics of chiral gauge theories: 
fermion number non-conservation due to chiral anomaly\cite{'tHooft:1976up, 'tHooft:1976fv}, 
various realizations of the gauge symmetry and global flavor 
symmetry\cite{Raby:1979my, Dimopoulos:1980hn},  
the existence of massless composite fermions suggested by 't Hooft's anomaly matching 
condition\cite{'tHooft:1979bh} and so on. 
Unfortunately,  very little is known so far about the actual behavior of chiral gauge theories beyond
perturbation theory.  It is desirable to develop a formulation to study  
the non-perturbative aspect of chiral gauge theories. 

Despite the well-known problem of the species 
doubling \cite{Karsten:1980wd,Nielsen:1980rz,Nielsen:1981xu,Friedan:1982nk}, 
lattice gauge theory can now provide
a framework for non-perturbative formulation of  chiral gauge theories. 
The clue to this development is 
the construction of local and gauge-covariant lattice Dirac operators 
satisfying the Ginsparg-Wilson relation\cite{Ginsparg:1981bj,
Neuberger:1997fp,Hasenfratz:1998ri,Neuberger:1998wv,
Hasenfratz:1998jp,Hernandez:1998et}.
By this relation, it is possible to realize an exact chiral symmetry on the lattice\cite{Luscher:1998pq}, 
without the species doubling problem. 
It is also possible to introduce Weyl fermions on the lattice and 
this opens the possibility
to formulate anomaly-free chiral lattice gauge theories\cite{Luscher:1998kn,Luscher:1998du,
Luscher:1999un,Luscher:1999mt,Luscher:2000hn,Suzuki:1999qw,Neuberger:2000wq,Adams:2000yi,
Suzuki:2000ii,Igarashi:2000zi,Luscher:2000zd,Kikukawa:2000kd,Kikukawa:2001jm}.
In the case of U(1) chiral gauge theories,  
L\"uscher\cite{Luscher:1998du}  proved rigorously that
it is possible to construct the fermion path-integral measure 
which depends smoothly on the gauge field  and
fulfills the fundamental requirements such as 
locality, gauge-invariance and lattice symmetries. 
Although it is believed that a chiral gauge theory is a difficult case for numerical simulations
because the effective action induced by Weyl fermions 
has a non-zero imaginary part,  it would be still interesting and even useful to 
develop a formulation of chiral lattice gauge theories by which one can work out 
fermionic observables numerically 
as the functions of link field with exact gauge invariance.\footnote{In the above formulation 
of U(1) chiral lattice gauge theories\cite{Luscher:1998du}, 
although the proof of the existence of the fermion measure is constructive,  the resulted formula 
of the fermion measure turns out to be rather complicated for the case of the finite-volume lattice. 
It also relies on the results obtained in the infinite lattice.  Therefore  it does not provide a formulation 
which is immediately usable for numerical applications. See 
\cite{Kadoh:2003ii,Kadoh:2004uu,Kadoh:2007} for a simplified formulation toward a practical implementation.}

In this article, we construct
the SU(2)$\times$U(1)  chiral gauge theory of the Glashow-Weinberg-Salam model\cite{Glashow:1961tr,Weinberg:1967tq,Salam:1968rm} on the lattice, keeping the exact gauge invariance.
As in the case of U(1) theories,
we first formulate the reconstruction theorem 
which asserts that
if there exists a set of local currents satisfying cetain properties, 
it is possible to reconstruct  the chiral 
fermion measure which depends smoothly on the gauge field  and 
fulfills the fundamental requirements such as locality\footnote{We adopt the generalized notion 
of locality on the lattice given in \cite{Hernandez:1998et, Luscher:1998kn, Luscher:1998du} 
for Dirac operators and composite fields.  
See also \cite{Kadoh:2003ii} for the case of the finite volume lattice.}, 
gauge-invariance and lattice symmetries.\footnote{The lattice symmetries mean
translations, rotations, reflections and charge conjugation.}
We then give a closed expression  of the local currents (the fermion measure term) 
for  the SU(2)$\times$U(1) chiral lattice gauge theory 
defined on the finite-volume lattice.  
Our construction covers all SU(2) topological sectors  
with vanishing U(1) magnetic fluxes.  
This formulation provides the first gauge-invariant and non-perturbative regularization 
of the electroweak theory, which would be usable in both perturbative and non-perturbative 
analyses. In particular, 
it  would be usable for a description of the baryon number 
non-conservation.\footnote{In the continuum theory, there is no winding number associated with 
the abelian gauge fields in four dimensions.  Therefore, we believe that the construction in the 
SU(2) topological sectors with vanishing U(1) magnetic fluxes would be sufficient for a 
description of the baryon number non-conservation.}

This article is organized as follows. 
In section~\ref{sec:GWS-on-lattice}, we introduce our lattice formulation of 
the Glashow-Weinberg-Salam model at the classical level.  
In section~\ref{sec:measure-reconstruction-theorem},  we define the  path-integral 
measure of chiral fermion fields and formulate the reconstruction theorem. 
In section~\ref{sec:measure-term-in-V}, 
we give an  explicit formula of  the local currents (the measure term) 
which fulfills all the required properties for the reconstruction theorem. 
In section~\ref{sec:measure-term-in-infinite-volume}, we discuss the measure term 
in the infinite volume limit. 
Section~\ref{sec:discussion} is devoted to discussions.

\section{The Glashow-Weinberg-Salam model on the lattice}
\label{sec:GWS-on-lattice}

In this section, we describe a construction of the Glashow-Weinberg-Salam model on the lattice
within the framework of chiral lattice gauge theories 
based on the lattice Dirac operator  satisfying the Ginsparg-Wilson 
relation \cite{Luscher:1998du,Luscher:1999un}.
We assume a local, gauge-covariant lattice Dirac operator $D$ which satisfies the Ginsparg-Wilson relation. An explicit example of such lattice Dirac operator is given by the overlap Dirac operator \cite{Neuberger:1997fp,Neuberger:1998wv}, 
which was derived from the overlap formalism 
\cite{
Narayanan:wx,Narayanan:sk,Narayanan:ss,Narayanan:1994gw,Narayanan:1993gq,
Neuberger:1999ry,Narayanan:1996cu,Huet:1996pw,
Narayanan:1997by,Kikukawa:1997qh,Neuberger:1998xn}.\footnote{
The overlap formula was derived from the five-dimensional approach of domain wall fermion proposed by Kaplan\cite{Kaplan:1992bt}.
In the vector-like formalism of domain wall fermion\cite{Shamir:1993zy, Furman:ky,
Blum:1996jf, Blum:1997mz}, 
the local low energy effective action of the chiral mode 
precisely reproduces the overlap Dirac 
operator \cite{Vranas:1997da,Neuberger:1997bg, Kikukawa:1999sy}.
}
In this case, our formulation is equivalent to the overlap formalism  
for chiral lattice gauge theories\footnote{
The overlap formalism gives a well-defined partition function of Weyl fermions on the lattice,
which nicely reproduces the fermion zero mode and the fermion-number
violating observables ('t Hooft vertices) \cite{Narayanan:1996kz,Kikukawa:1997md,Kikukawa:1997dv}. 
The gauge-invariant construction by L\"uscher \cite{Luscher:1998du} 
provides a procedure to fix the ambiguity of  the complex phase of the overlap formula
 in a gauge-invariant manner for anomaly-free U(1) chiral gauge theories.
}
or  the domain wall fermion approach \cite{Kikukawa:2001mw, Aoyama:1999hg}.
See \cite{Creutz:1996xc} for the attempt to construct the standard model  
in the domain wall fermion approach combined with the construction by Eichten and Preskill 
\cite{Eichten:1985ft}.\footnote{See 
also \cite{Bhattacharya:2006dc,Giedt:2007qg,Poppitz:2007tu,Gerhold:2007yb,Gerhold:2007gx,Kikukawa:2007im} 
for the recent attempt to construct chiral gauge 
theories using
mirror Ginsparg-Wilson fermions with gauge- and chiral-invariant Yukawa couplings to 
the extra bosonic degrees of freedom, which may be identified with the 
Higgs field or Wess-Zumino scalar field,  and for related works.}

\subsection{SU(2)$\times$U(1) Gauge fields}

We consider the four-dimensional lattice of the finite size $L$ and choose lattice units, 
\begin{equation}
\Gamma =
 \left\{x=(x_1, x_2, x_3, x_4)  \in \mathbb{Z}^4 \, \vert \, \, 0 \le x_\mu < L \,  (\mu = 1,2,3,4) \right\} . 
\end{equation}
Adopting the  compact formulation for U(1) lattice gauge theory, 
the SU(2) and U(1) gauge fields on $\Gamma$ may be represented through 
periodic link fields on the infinite lattice: 
\begin{eqnarray}
&& U^{(1)}(x,\mu) \, \in \text{U(1)},    \qquad  x  \in \mathbb{Z}^4,  \\
&& U^{(1)}(x+L\hat \nu, \mu) = U^{(1)}(x,\mu) \quad \text{for all }  \mu,\nu, 
\end{eqnarray}
and 
\begin{eqnarray}
&& U^{(2)}(x,\mu) \, \in \text{SU(2)},    \qquad  x  \in \mathbb{Z}^4,  \\
&& U^{(2)}(x+L\hat \nu, \mu) = U^{(2)}(x,\mu) \quad \text{for all }  \mu,\nu. 
\end{eqnarray}
%
We require the so-called admissibility condition on the gauge fields, 
\begin{eqnarray}
&& 
\vert F_{\mu\nu}(x) \vert  < \epsilon_1, 
\qquad F_{\mu\nu}(x) \equiv \frac{1}{i} {\rm ln} P^{(1)}(x,\mu,\nu) \in (-\pi, \pi] ,  \\
&&
\|1-P^{(2)}(x,\mu,\nu)  \|  < \epsilon_2 , 
\end{eqnarray}
for all $x$, $\mu, \nu$, where the plaquette variables are defined by 
\begin{eqnarray}
P^{(i)}(x,\mu,\nu)&=&
U^{(i)}(x,\mu)U^{(i)}(x+\hat\mu,\nu)U^{(i)}(x+\hat\nu,\mu)^{-1}U^{(i)}(x,\nu)^{-1}  \quad (i=1,2) .  
\end{eqnarray}
This condition ensures that 
the overlap Dirac operator\cite{Neuberger:1997fp,Neuberger:1998wv} is a
smooth and local function of  the gauge field 
if $ (Y\epsilon_1) <1/30$ and  $ \epsilon_2 <1/30$, where 
$Y$ is the hyper-charge of the fermion on which the overlap 
Dirac operator acts \cite{Hernandez:1998et}.

To impose the admissibility condition dynamically, we adopt the following action for the gauge fields:
\begin{eqnarray}
S_G &=& \frac{1}{g_2^2} \sum_{x\in \Gamma} \sum_{\mu,\nu}
 {\rm tr}\{ 1-P^{(2)}(x,\mu,\nu) \}  
  \left[ 1 -   {\rm tr}\{ 1-P^{(2)}(x,\mu,\nu) \}/ \epsilon_2^2 \right]^{-1}  \nonumber\\
&&+  \frac{1}{4g_1^2} \sum_{x\in \Gamma} \sum_{\mu,\nu}
  \left[ F_{\mu\nu}(x) \right]^2 
  \left\{ 1 -  \left[ F_{\mu\nu}(x) \right]^2 / \epsilon_1^2 \right\}^{-1} . 
\end{eqnarray}

\subsection{Quarks and Leptons}
Right- and left- handed Weyl fermions are introduced  on the lattice based on the Ginsparg-Wilson relation.
Let us first consider  a generic gauge group $G$ and a Dirac field $\psi(x)$ 
coupled to the  gauge field $U(x,\mu)$ in a certain representaion $R$ of $G$. 
Then we assume a local, gauge-covariant lattice Dirac operator $D_L$ which acts on $\psi(x)$ and satisfies the Ginsparg-Wilson relation, 
\begin{equation}
\label{eq:GW-rel}
\gamma_5 D_L + D_L  \gamma_5  = 2 D_L \gamma_5 D_L.  
\end{equation}
 The kernel of the lattice Dirac operator in finite volume, $D_L$,  may be represented 
through the kernel of the lattice Dirac operator in infinite volume, $D$, as follows:
\begin{equation}
\label{eq:DL-in-D-infity}
D_L(x,y) = D(x,y) + \sum_{n \in \mathbb{Z}^4, n \not = 0} D(x,y+n L) ,
\end{equation}
where $D(x,y)$ is defined with a periodic link field in  infinite volume. We assume that 
$D(x,y)$ posseses the locality property given by 
\begin{equation}
\label{eq:locality-D}
\| D(x,y) \|  \le C ( 1+\| x-y \|^p ) \, {\rm e}^{-\| x-y \| / \varrho}
\end{equation}
for some constants $\varrho > 0$, $C >0$ , $p \ge 0$, 
where  $\varrho$ is the localization range of the lattice Dirac operator.

Given such a lattice Dirac operator $D_L$, one can  introduce a chiral operator as 
\begin{equation}
\label{eq:hat-gamma5} 
 \hat \gamma_5 \equiv  \gamma_5(1- 2 D_L), \qquad (\hat \gamma_5 )^2 = \mathbb{I}. 
\end{equation}
Then, the right- and left-handed Weyl fermions in the representaion $R$ of $G$ can be defined by the eigenstates of the chiral 
operator $\hat \gamma_5$ (and $\gamma_5$ for the anti-fields). Namely, 
\begin{equation}
\psi_\pm(x) = \hat P_\pm \psi(x) ,  \qquad
\bar \psi_\pm(x) = \bar \psi(x) P_\mp ,    
\end{equation}
where $\hat P_\pm$ and $P_\pm$ are the chiral projection operators given by 
\begin{equation}
\hat P_{\pm} = \left( \frac{1\pm \hat \gamma_5}{2} \right) , \quad
 P_{\pm} = \left( \frac{1\pm  \gamma_5}{2} \right) . 
 \end{equation}
 
Now we consider quarks and leptons in the Glashow-Weinberg-Salam model. For simplicity, we consider the first family.  
We adopt the convention for the normalization of the hyper-charges such that the Nishijima-Gell-Mann relation reads  $Q=T_3+\frac{1}{6} Y$.
To describe the left-handed quarks and leptons, 
which are SU(2) doublets,  
we introduce a left-handed fermion $\psi_{-}(x)$ with the index $\alpha (=1,\cdots,4)$, 
each component of  which couples to the SU(2)$\times$U(1) gauge field,  
$U^{(2)}(x,\mu)\otimes \{U^{(1)}(x,\mu)\}^{Y_\alpha}$, with the hyper-charge 
$Y_\alpha$ ( $Y_{1,2,3}=1$ and $Y_4=-3$). Namely, 
\begin{eqnarray}
\psi_-(x) &=& {}^t \left( q_-^1(x), q_-^2(x), q_-^3(x), l_-(x) \right).
\end{eqnarray}
Similarly, to describe the right-handed quarks and leptons, 
which are SU(2) singlets, 
we introduce a right-handed fermion $\psi_{+}(x)$ with the index 
$\beta (=1,\cdots,8)$, each component of  which couples to the U(1) gauge field,  
$\{U^{(1)}(x,\mu)\}^{Y_\beta}$, with the hyper-charge $Y_\beta$ 
($Y_{1,3,5}=4$, $Y_{2,4,6}=-2$, $Y_7=0$ and $Y_8=-6$).  Namely, 
\begin{eqnarray}
\psi_+(x) &=& 
{}^t \left( u_+^1(x), d_+^1(x),  u_+^2(x), d_+^2(x), u_+^3(x), d_+^3(x), \nu_+(x), e_+(x) \right).
\end{eqnarray}
Then the action of quarks and leptons is given  by 
\begin{equation}
S_F = \sum_{x\in\Gamma} \bar \psi_- (x) D_L \psi_-(x) +
           \sum_{x\in\Gamma} \bar \psi_+ (x) D_L \psi_+(x) . 
\end{equation}

\subsection{Higgs field and its Yukawa-couplings to quarks and leptons}

Higgs field is a SU(2) doublet with the hyper-charge $Y_h=+6$. 
The action of the Higgs field may be given by
\begin{equation}
S_H = \sum_x \left[ \sum_{\nu} \left(\nabla_\nu \phi(x)\right)^\dagger \nabla_\nu \phi(x) 
+ \frac{\lambda}{2} \left(  \phi(x)^\dagger \phi(x)  - v^2 \right)^2 \right], 
\end{equation}
where $\phi(x)$ couples to the gauge field $U^{(2)}(x,\mu)\otimes \{U^{(1)}(x,\mu)\}^{Y_h}$
and $\nabla_\nu$ is the SU(2)$\times$ U(1) gauge-covariant difference operator. 
Yukawa couplings of the Higgs field to the quarks and leptons may also be introduced 
as follows\footnote{One may add the Dirac-type mass term for the neutrino, 
$\sum_x \{ y_\nu \, \bar l_-(x) \tilde \phi(x) \nu_+(x) \ + y_\nu^\ast \, \bar \nu_+(x) \tilde 
\phi(x)^\dagger l_-(x)\}$.}: 
\begin{eqnarray}
S_Y &=& \sum_x \left[ 
   y_u \, \bar q_-^i(x) \tilde \phi(x) u_+^i(x) + y_u^\ast \, \bar u_+^i(x) \tilde \phi(x)^\dagger q_-^i(x) 
                               \right.\nonumber\\
 && \qquad 
 +y_d \, \bar q_-^i(x) \phi(x) d_+^i(x) + y_d^\ast  \, \bar d_+^i(x) \phi(x)^\dagger q_-^i(x)
 \nonumber\\
&& \qquad \ \left. 
+ y_l \, \bar l_-(x) \phi(x) e_+(x) \ + y_l^\ast \, \bar e_+(x) \phi(x)^\dagger l_-(x)  \right] , 
 \end{eqnarray}
 where $\tilde \phi(x)$ is the SU(2) conjugate of $\phi(x)$. 
 
Thus the total lattice action, 
\begin{equation}
S=S_G +S_F + S_H + S_Y, 
\end{equation}
defines a classical theory of the Glashow-Weinberg-Salam model on the lattice with the first-family 
quarks and leptons. 
In this action, locality, 
gauge-invariance and lattice symmetries such as translations and rotations are manifest. CP symmetry, however, is not manifest even with the real Yukawa couplings. But it is possible to show that at the quantum level both the partition function and the on-shell amplitudes respect the 
CP symmetry \cite{Suzuki:2000ku,Fujikawa:2002vj,Fujikawa:2002up,Giusti:2002rx}. 
With the three families, then, the breaking of CP symmetry comes from the 
Kobayashi-Maskawa phase\cite{Kobayashi:1973fv} as in the continuum theory. 

\subsection{Topology of the  SU(2)$\times$U(1) gauge fields}
The admissibility condition ensures that 
the overlap Dirac operator\cite{Neuberger:1997fp,Neuberger:1998wv}
is a smooth and local function of  the gauge field \cite{Hernandez:1998et}. 
Then,  through the lattice Dirac operator $D_L$, 
it is possible to define a topological 
charge of the gauge fields \cite{Narayanan:sk,Narayanan:ss,Narayanan:1993gq,
Hasenfratz:1998ri,Luscher:1998pq}: 
for the admissible SU(2) and U(1) gauge fields, one has
\begin{equation}
Q^{(i)} = \left. {\rm Tr}\gamma_5(1- D_L) \right\vert_{U=U^{(i)}} 
        =\sum_{x  \in \Gamma}  
        \left. {\rm tr}\left\{ \gamma_5(1- D_L) \right\}(x,x) \right\vert_{U=U^{(i)}} \quad (i=1,2), 
\end{equation}
where $D_L(x,y)$ is the kernel of the lattice Dirac operator $D_L$. 
For $0 < \epsilon_1 < \pi/3$, the admissible U(1) gauge fields can also be classified by the magnetic 
fluxes, 
\begin{equation}
m_{\mu\nu} 
=
\frac{1}{2\pi}\sum_{s,t=0}^{L-1}F_{\mu\nu}(x+s\hat\mu+t\hat\nu) , 
\end{equation}
which are integers independent of $x$.  
$m_{\mu\nu}$ is related to $Q^{(1)}$ by $Q^{(1)} = (1/2)\sum_{\mu\nu} m_{\mu\nu}^2$ \cite{Fujiwara:1999fj}. 
Then the admissible SU(2) and U(1) gauge fields can be classified by the topological numbers $Q_2$ and $m_{\mu\nu}$, respectively.\footnote{Strictly speaking, the complete topological classification of the space of
admissible SU(2) gauge fields is not known yet.  However, as we will see,  our construction
 is valid for any SU(2) topological sectors, as long as the U(1) magnetic flux vanishes identically. } 
We denote the space of  the admissible SU(2) gauge fields with a given topological charge 
$Q^{(2)}$ by $\mathfrak{U}^{(2)}[Q]$ and the space of the admissible U(1) gauge fields with a given magnetic fluxes $m_{\mu\nu}$ by $\mathfrak{U}^{(1)}[m]$. 

\section{Path-integral measure of the lattice Glashow-Weinberg-Salam model}
\label{sec:measure-reconstruction-theorem}

In this section, we 
consider a construction of the path-integral measure of the 
quarks and leptons in the lattice Glashow-Weinberg-Salam model.\footnote{The path-integral measure of the SU(2)$\times$U(1) gauge fields and 
Higgs field may be defined as usual.}  
We will show that, as in the case of the U(1) chiral gauge theories  \cite{Luscher:1998du}, 
it is possible to formulate a reconstruction theorem of the fermion measure 
for the topological sectors of the admissible SU(2)$\times$U(1) gauge fields with vanishing U(1) magnetic fluxes.  This reconstruction theorem asserts that
if there exist local currents which satisfy cetain properties, 
it is possible to reconstruct  the 
fermion measure 
which depends smoothly on the gauge field  and 
fulfills the fundamental requirements such as 
locality, 
gauge-invariance and lattice symmetries.

\subsection{Path-integral measure of Quarks and Leptons}
\label{subsec:properties_of_measure_term}

The path-integral measure of quark fields and lepton fields may be defined by
the Grassmann integrations, 
\begin{equation}
{\cal D}[\psi_+] {\cal D}[\bar \psi_+] {\cal D}[\psi_-] {\cal D}[\bar \psi_-]  
= \prod_j d b_j  \prod_k d \bar b_k \prod_j d c_j  \prod_k d \bar c_k  ,   
\end{equation}
where $\{ b_j , \bar b_k\}$ and $\{c_j ,  \bar c_k \}$ are the grassman coefficients
in the expansion of the chiral fields, 
\begin{equation}
\psi_+(x) = \sum_j  u_j(x) b_j , \quad  \bar \psi_+(x) = \sum_k \bar b_k \bar u_k(x) , 
\end{equation}
\begin{equation}
\psi_-(x) = \sum_j  v_j(x) c_j , \quad  \bar \psi_-(x) = \sum_k \bar c_k \bar v_k(x) , 
\end{equation}
 in terms of the chiral (orthonormal) bases defined by 
\begin{equation}
\hat P_{+} u_j(x) = u_j(x) ,    \quad 
\bar u_k (x) P_{-}  = \bar u_k (x) .  
\end{equation}
\begin{equation}
\hat P_{-} v_j(x) = v_j(x) ,    \quad 
\bar v_k (x) P_{+}  = \bar v_k (x) .  
\end{equation}
Since the projection operators $\hat P_{\pm} $ depend on the gauge fields through $D$, 
the fermion measure also depends on the gauge fields.  

This gauge field dependence can be examined explicitly by considering the effective action
induced by the quarks and leptons, 
\begin{equation}
\Gamma_{\rm eff} 
= \ln \left[  \det (\bar v_k D_L v_j) \, \det (\bar u_k D_L u_j) \right].
\end{equation}
With respect to the variation of the gauge fields,
\begin{eqnarray}
&& \delta_\eta U^{(1)}(x,\mu) = i \eta^{(1)}_\mu(x) U^{(1)}(x,\mu), \\
\label{eq:variation-su2}
&&\delta_\eta U^{(2)}(x,\mu) = i \eta^{(2)}_\mu(x)  U^{(2)}(x,\mu), \qquad 
(\eta^{(2)}_\mu(x)  \equiv \eta^{a}_\mu(x) T^a) , 
\end{eqnarray}
the variation of the effective action $\Gamma_{\rm eff}$ is evaluated as 
\begin{eqnarray}
\delta_\eta \Gamma_{\rm eff} &=& {\rm Tr} \left\{ \delta_\eta D_L \hat P_- D_L^{-1} P_+ \right\} 
                                      +{\rm Tr} \left\{ \delta_\eta D_L \hat P_+ D_L^{-1} P_- \right\}  
                                      +  \sum_j  ( v_j , \delta_\eta v_j ) +   \sum_j  ( u_j , \delta_\eta u_j ). 
                                      \nonumber\\
\end{eqnarray}
In particular, for the gauge transformations
\begin{eqnarray}
&& \eta^{(1)}_\mu(x) = - \partial_\mu \omega(x), \\
\label{eq:gauge-transformation-su2}
&& \eta^{(2)}_\mu(x)= - [\nabla_\mu \omega]^a(x) T^a, 
\end{eqnarray}
it is given as
\begin{eqnarray}
\delta_\omega \Gamma_{\rm eff} &=&
 i \sum_{x \in \Gamma} \omega(x) 
        \left[{\rm tr}\{ Y_- \gamma_5(1-D_L)(x,x) \} -{\rm tr}\{ Y_+ \gamma_5(1-D_L)(x,x) \}\right]
 \nonumber\\
&+&
i \sum_{x \in \Gamma} \omega^a(x) \, {\rm tr}\{ T^a \gamma_5(1-D_L)(x,x) \}
+ \sum_j  ( v_j , \delta_\omega v_j ) +  \sum_j  ( u_j , \delta_\omega u_j ), 
\end{eqnarray}
where $Y_-=\text{diag}(1,1,1,-3)$ and $Y_+=\text{diag}(4,-2, \cdots, 0, -6)$.\footnote{Throughout this paper, 
${\text Tr}$ stands for the trace over the lattice index $x \, (\in \Gamma)$, 
the flavor indices $\alpha$,  $\beta$ and the spinor index.   
${\text tr}$ stands for the trace over the flavor and/or spinor indices only.}   

In this gauge-field dependence
of the fermion measure, there is  an ambiguity by a pure phase factor, because
any unitary transformations of the bases, 
\begin{equation}
\tilde u_j(x) = \sum_l u_l (x) \left(  {\cal Q_+}^{-1} \right)_{lj},  \qquad 
\tilde b_j = \sum_l   \left({\cal Q}_+\right)_{jl}  b_l , 
\end{equation}
\begin{equation}
\tilde v_j(x) = \sum_l v_l (x) \left(  {\cal Q_-}^{-1} \right)_{lj},  \qquad 
\tilde c_j = \sum_l   \left({\cal Q}_-\right)_{jl}  c_l , 
\end{equation}
induces a change of the measure by the pure phase factor $\det {\cal Q_+} \cdot \det {\cal Q_-}$.
This ambiguity should be fixed
so that
the measure fulfills the fundamental requirements such as 
locality, 
gauge-invariance, integrability and lattice symmetries. 


\subsection{Gauge anomaly cancellations in the lattice Glashow-Weinberg-Salam model} 

We next examine the gauge anomaly cancellations in the lattice Glashow-Weinberg-Salam model. 

\subsubsection{Pseudo reality of SU(2) and the absence of SU(2)$^3$ gauge anomaly}

We first consider the case where the U(1) link field is trivial. 
In  the topological sectors with vanishing U(1) magnetic flux, 
$\mathfrak{U}^{(2)}[Q] \otimes \mathfrak{U}^{(1)}[0]$, 
any admissible U(1) link field can be continuously deformed to the trivial configuration,
$U^{(1)}(x,\mu) = 1$.  
In this limit,  
only the SU(2) gauge field couples to the left-handed fermion $\psi_-(x)$, which now consists of 
four degenerate SU(2) doublets. 
By noting the pseudo reality of SU(2), 
\begin{equation}
U^{(2)}(x,\mu)^\ast = (i \sigma_2)  \, U^{(2)}(x,\mu) \, (i \sigma_2 )^{-1}, 
\end{equation}
and the charge- and $\gamma_5$-conjugation properties of the lattice Dirac operator, 
\begin{equation}
D_L[ {U^{(2)}}^\ast ] = C^{-1}  \{ D_L[U^{(2)}]\}^T   C, \quad 
D_L[ {U^{(2)}}]^\dagger = \gamma_5 D_L[U^{(2)}]  \gamma_5, 
\end{equation}
where $C$ is the charge conjugation matrix satisfying $C \gamma_\mu C^{-1} = - \gamma_\mu^T$, 
one can infer that
\begin{equation}
D_L[ {U^{(2)}}] 
= (\gamma_5  C^{-1}\otimes  i\sigma_2)  
  \{ D_L[U^{(2)}]\}^\ast  ( C \gamma_5 \otimes (i\sigma_2)^{-1} ) . 
\end{equation}
Then one may choose the basis vectors of the left-handed fermion 
$\psi_-(x)$$={}^t\big($$q_-^1(x)$, $q_-^2(x)$, $q_-^3(x)$, $l_-(x)$$\big)$ 
for any given SU(2) gauge field $U^{(2)}(x,\mu) \in \mathfrak{U}^{(2)}[Q] $ as follows:
\begin{eqnarray}
\label{eq:four-doublets-basis-q1}
q_-^1(x) &=& \sum_j w_j(x) c_j^1, \\
\label{eq:four-doublets-basis-q2}
q_-^2(x) &=& \sum_j \left( \gamma_5 C^{-1} \otimes  i \sigma_2 \right) \, \left[ w_j(x) \right]^\ast c_j^2, \\
\label{eq:four-doublets-basis-q3}
q_-^3(x) &=& \sum_j w_j(x) c_j^3, \\
\label{eq:four-doublets-basis-l}
l_-(x) &=& \sum_j  \left( \gamma_5 C^{-1} \otimes  i \sigma_2 \right) \, \left[ w_j(x) \right]^\ast c_j^4, 
\end{eqnarray}
where $\{w_j(x)\}$ is an arbitrarily chosen basis for a single left-handed SU(2) doublet.  
With this choice of the basis,  one can infer that the measure term vanishes identically and 
therefore the fermion measure is manifestly invariant under the SU(2) gauge transformation, 
eqs.~(\ref{eq:variation-su2}) and (\ref{eq:gauge-transformation-su2}).

\subsubsection{Cancellations of SU(2)$^2$$\times$U(1) and U(1)$^3$ gauge anomalies} 
\label{subsec:gauge-anomaly-cancellation-u1}

When the U(1) link field is non-trivial in generic topological sectors, 
$\mathfrak{U}^{(2)}[Q] \otimes \mathfrak{U}^{(1)}[m]$, 
the  U(1) part of the gauge anomaly 
is given by %
\begin{equation}
q_L^{(1)}(x) =  {\rm tr}\{ Y_- \gamma_5(1-D_L)(x,x) \} -{\rm tr}\{ Y_+ \gamma_5(1-D_L)(x,x) \}, 
\end{equation}
where $D_L(x,y)$ is the finite-volume kernel of the lattice Dirac operator.   
It is topological in the sense that
\begin{equation}
\sum_{x \in \Gamma} q_L^{(1)}(x) = \text{integer} , \qquad \qquad
\sum_{x \in \Gamma} \delta_\eta  q_L^{(1)}(x) = 0. 
\end{equation}
Then the following lemma holds true concerning the cancellations 
of SU(2)$^2$$\times$U(1) and U(1)$^3$ gauge anomalies:

\vspace{1em}
\noindent{\bf Lemma 1} \ \ 
{\sl In the lattice Glashow-Weinberg-Salam model, 
the U(1) gauge anomaly has the following form in sufficiently large volume $L^4$: 
\begin{eqnarray}
\label{eq:gauge-anomaly-u1-reslut1}
q_L^{(1)}(x) &=&   
\left.  {\rm tr}\{ Y_- \gamma_5(1-D_L)(x,x) \} \right\vert_{U=U^{(2)}} \nonumber\\
&+& \big( {\rm tr}\{ Y_-^3\} - {\rm tr}\{ Y_+^3\} \big)
\, \gamma \,   \epsilon_{\mu\nu\lambda\rho}F_{\mu\nu}(x) F_{\lambda\rho}(x+\hat\mu+\hat\nu) 
\nonumber\\
&+& \partial_\mu^\ast  k_\mu(x) , 
\end{eqnarray} 
where 
$\gamma$ is a constant independent of the gauge fields
and  $k_\mu(x)$ is a local, gauge-invariant  current which
can be constructed so that it transforms as an axial vector current under the lattice symmetries. 
Moreover,  since the hyper-charges of a single family of quarks and leptons satisfy the anomaly cancellation conditions,
\begin{eqnarray}
&&  {\rm tr} \{ Y_- \} = 0, \\ 
&&  {\rm tr}\{ Y_-^3\} - {\rm tr}\{ Y_+^3\} = 0 , 
\end{eqnarray}
the cohomologically non-trivial part of the gauge anomaly cancels exactly at a finite lattice spacing 
and the total U(1) gauge anomaly is cohomologically trivial:
\begin{equation}
q_L^{(1)}(x) 
= \partial_\mu^\ast k_\mu(x) . 
\end{equation}
}

\vspace{1em}
\noindent{\it Proof} : 
Given the topological property of the U(1) gauge anomaly, 
it is possible to apply the cohomology-analysis developed for 
the U(1) case \cite{Luscher:1998kn,Fujiwara:1999fi,Fujiwara:1999fj,Igarashi:2002zz} to 
the SU(2)$\times$U(1) case by regarding the SU(2) gauge field $U^{(2)}(x,\mu)$ as a 
background.\footnote{This trick was first used
to show the gauge anomaly cancellation in the lattice
Glashow-Weinberg-Salam model \cite{Kikukawa:2000kd}  in the 4+2 dimensional approach to the cohomological analysis of non-abelian gauge 
anomalies \cite{Luscher:1999un,Adams:2000yi,Alvarez-Gaume:1983cs}. }
The result is given by the following expression: 
\begin{eqnarray}
\label{eq:gauge-anomaly-u1-reslut0}
q_L^{(1)}(x) &=&   
\left.  {\rm tr}\{ Y_- \gamma_5(1-D_L)(x,x) \} \right\vert_{U=U^{(2)}} \nonumber\\
&+& \beta_{\mu\nu}(x) F_{\mu\nu}(x)  \nonumber\\
&+& \big( {\rm tr}\{ Y_-^3\} - {\rm tr}\{ Y_+^3\} \big)
\, \gamma \,   \epsilon_{\mu\nu\lambda\rho}F_{\mu\nu}(x) F_{\lambda\rho}(x+\hat\mu+\hat\nu) 
\nonumber\\
&+& \partial_\mu^\ast  k_\mu(x) , 
\end{eqnarray} 
where 
$\gamma$ is a constant independent of the gauge fields,  
which takes the value 
$\gamma = \frac{1}{32 \pi^2}$ for the overlap Dirac operator \cite{Kikukawa:1998pd}, 
$\beta_{\mu\nu}(x)$ is a tensor field satisfying 
$\partial_\mu^\ast \beta^{}_{\mu\nu}(x)=0$ which depends only on the SU(2) gauge field 
and  $k_\mu(x)$ is a local, gauge-invariant  current which
can be constructed so that it transforms as the axial vector current under the lattice symmetries. 
Moreover, taking into account the pseudo-scalar nature of $q_L^{(1)}(x)$ under the charge conjugation and the pseudo reality of SU(2),  one has
\begin{equation}
\left. q_L^{(1)}(x) \right\vert_{U=U^{(2)}, {U^{(1)}}^\ast} 
= \left. q_L^{(1)}(x) \right\vert_{U=U^{(2)}, {U^{(1)}}}, 
\end{equation}
which immediately implies that the second term in the r.h.s. of 
eq.~(\ref{eq:gauge-anomaly-u1-reslut0}) can be included into 
the total-divergence term as 
\begin{equation}
2 \, \beta_{\mu\nu}(x) F_{\mu\nu}(x) = 
 \partial_\mu^\ast \, \left[ \left. k_\mu(x) \right\vert_{U=U^{(2)}, {U^{(1)}}^\ast} -
                                           \left. k_\mu(x) \right\vert_{U=U^{(2)}, {U^{(1)}} } \right]. 
\end{equation}

\vspace{1.0em}
We emphasize that this is the result of the U(1) gauge anomaly  {\it in finite volume},  which is obtained by combining the result in the infinite lattice \cite{Luscher:1998kn,Fujiwara:1999fi,Fujiwara:1999fj} with
the use of the trick to regard the SU(2) gauge field $U^{(2)}(x,\mu)$ as a background \cite{Kikukawa:2000kd}, and the result of the analysis of the finite volume correction \cite{Igarashi:2002zz}.  
See also \cite{Kadoh:2003ii}. 
In fact,  the local, gauge-invariant current $k_\mu(x)$ may be decomposed as
\begin{equation}
k_\mu(x) = \bar k_\mu(x) + \Delta k_\mu(x), 
\end{equation}
where $\bar k_\mu(x) $ and $\Delta k_\mu(x)$ satisfy the anomalous conservation laws,
\begin{eqnarray}
\partial_\mu^\ast k^{}_\mu(x) &=& 
{\rm tr}\{ Y_- \gamma_5(1-D)(x,x) \} -{\rm tr}\{ Y_+ \gamma_5(1-D)(x,x) \}  \nonumber\\
&\equiv& q^{(1)}(x), \\
&&\nonumber\\
\partial_\mu^\ast \Delta k_\mu(x) &=& 
\sum_{n \in \mathbb{Z}^4, n\not = 0} \left[  {\rm tr}\{ Y_- \gamma_5(1-D)(x,x+Ln) \} 
-{\rm tr}\{ Y_+ \gamma_5(1-D)(x,x+Ln) \} \right] \nonumber\\
& \equiv & r(x) , 
\end{eqnarray}
respectively.  $\bar k_\mu(x)$ is obtained as the solution of 
the cohomology-analysis
\cite{Luscher:1998kn,Fujiwara:1999fi,Fujiwara:1999fj} applied  to 
$q^{(1)}(x)$ in infinite volume, 
while $\Delta k_\mu(x)$ is the result of the analysis of the finite volume 
correction \cite{Igarashi:2002zz} applied  to $r(x)$, both in the use of the trick to 
regard the SU(2) gauge field $U^{(2)}(x,\mu)$ as a background \cite{Kikukawa:2000kd}. 
One can infer from eq.~(\ref{eq:locality-D}) that 
\begin{eqnarray}
\label{eq:bound-delta-kmu}
&& \left| \Delta k_\mu(x) \right| \le  C_1  \, {\rm e}^{-L/\varrho} 
\end{eqnarray}
for a constant $C_1 > 0$ \cite{Igarashi:2002zz}.
This result should be compared with the result 
obtained from the 4+2 dimensional approach to the cohomological analysis of non-abelian gauge anomalies \cite{Kikukawa:2000kd}, where only the solution in the infinite volume limit has been 
obtained so far. 

\subsubsection{Issue related to SU(2) global anomaly}

When the U(1) link field is trivial in  the topological sectors with vanishing U(1) magnetic flux, 
$\mathfrak{U}^{(2)}[Q] \otimes \mathfrak{U}^{(1)}[0]$, 
one can construct the fermion measure which is invariant under the SU(2) gauge transformation, 
eqs.~(\ref{eq:variation-su2}) and (\ref{eq:gauge-transformation-su2}).  However, there remains 
the issue related to  SU(2) global anomaly\cite{Neuberger:1998rn,Bar:2002sa,Bar:2000qa}. 
In the following sections, we will establish rigorously  that  
the lattice counterpart of the SU(2) global 
anomaly \cite{Neuberger:1998rn,Bar:2002sa,Bar:2000qa} is absent 
in the topological sectors with vanishing U(1) magnetic flux, 
$\mathfrak{U}^{(2)}[Q] \otimes \mathfrak{U}^{(1)}[0]$.

\subsection{Reconstruction theorem of the fermion measure}

We now formulate the reconstruction theorem of the fermion measure
in the lattice Glashow-Weinberg-Salam model. 
The properties of the fermion measure can be characterized by the so-called measure term
which is given in terms of the chiral basis and its variation with respect to the gauge fields as
\begin{equation}
\label{eq:measure-term}
\mathfrak{L}_\eta =  i \sum_j  ( v_j , \delta_\eta v_j ) +  i \sum_j  ( u_j , \delta_\eta u_j ). 
\end{equation}
Similar to the case of U(1) chiral lattice gauge theories \cite{Luscher:1998du}, 
one can establish the following theorem. 

\vspace{1em}
\noindent
{\bf Theorem:} {\sl In  the topological sectors with vanishing U(1) magnetic flux, 
$\mathfrak{U}^{(2)}[Q] \otimes \mathfrak{U}^{(1)}[0]$, if there exist local 
currents $j^a_\mu(x)$$(a=1,2,3)$, $j_\mu(x)$ 
which satisfy the following four properties, 
it is possible 
to reconstruct  the fermion measure (the bases  $\{ u_j(x) \}$,  $\{ v_j(x) \}$) 
which depends smoothly on the gauge fields  and
fulfills the fundamental requirements such as 
locality, 
gauge-invariance, integrability and lattice symmetries: 
\begin{enumerate}
\item  $j^a_\mu(x)$, $j_\mu(x)$ are defined for all  admissible SU(2)$\times$U(1) gauge fields in the given topological sectors and depends smoothly on the link variables.

\item $j^a_\mu(x)$ and $j_\mu(x)$ are gauge-covariant and -invariant, respectively
 and both transform as axial vector currents under the lattice symmetries. 

\item The linear functional 
$\mathfrak{L}_\eta= \sum_{x\in \Gamma} \{ \eta^a_\mu(x) j^a_\mu(x)+ \eta_\mu(x) j_\mu(x) \}$
is a solution of the integrability condition
\begin{equation}
\label{eq:integrability-condition}
\delta_\eta \mathfrak{L}_\zeta - \delta_\zeta \mathfrak{L}_\eta +\mathfrak{L}_{[\eta,\zeta]}
=
i {\rm Tr} \left\{ \hat P_- [ \delta_\eta \hat P_-, \delta_\zeta \hat P_- ] \right\} 
+ i {\rm Tr} \left\{ \hat P_+ [ \delta_\eta \hat P_+, \delta_\zeta \hat P_+ ] \right\} 
\end{equation}
for all periodic variations $\eta^a_\mu(x), \eta_\mu(x)$ and $\zeta^a_\mu(x), \zeta_\mu(x)$.

\item The anomalous conservation laws hold: 
\begin{eqnarray}
\label{eq:anomalous-conservation-laws-su2}
\{\nabla_\mu^\ast j_\mu\}^a (x) 
&=&    {\rm tr}\{ T^a \gamma_5(1-D)(x,x) \},  \\
\label{eq:anomalous-conservation-laws-u1}
\partial_\mu^\ast j_\mu(x) 
&=&   {\rm tr}\{ Y_- \gamma_5(1-D_L)(x,x) \} -{\rm tr}\{ Y_+ \gamma_5(1-D_L)(x,x) \}, 
\end{eqnarray}
where $Y_-=\text{diag}(1,1,1,-3)$ and $Y_+=\text{diag}(4,-2, \cdots, 0, -6)$.  
\end{enumerate}
}

\vspace{1em}
A comment is in order about the topological aspects of the reconstrtuction theorem. 
It is possible,  as discussed in \cite{Luscher:1998du}, to associate a U(1) bundle with the 
fermion measure. In this point of view, 
the measure term,  $\mathfrak{L}_\eta$ defined by eq.~(\ref{eq:measure-term}), 
can be regarded as the connection of the U(1) bundle,  and 
the quantity which appears in the r.h.s. of the  integrability condition 
eq.~(\ref{eq:integrability-condition}),  
\begin{equation}
\label{eq:curvature-term}
\mathfrak{C}_{\eta\zeta} \equiv 
 i {\rm Tr} \left\{ \hat P_- [ \delta_\eta \hat P_-, \delta_\zeta \hat P_- ] \right\}  
+ i {\rm Tr} \left\{ \hat P_+ [ \delta_\eta \hat P_+, \delta_\zeta \hat P_+ ] \right\}
\end{equation}
is nothing but the curvature of the connection, 
\begin{equation}
\mathfrak{C}_{\eta\zeta} = \delta_\eta \mathfrak{L}_\zeta -  \delta_\zeta \mathfrak{L}_\eta 
  + \mathfrak{L}_{[\eta,\zeta]}. 
\end{equation} 
It is known that the integration of the curvature of a U(1) bundle 
over any two-dimensional closed surface in the base manifold
takes value of  the multiples of $2\pi$. If one parametrize a two-dimensional closed surface 
in the space of the admissible U(1) gauge fields 
by $s, t \in [ 0,2\pi]$, then one has  
\begin{equation}
\int_0^{2\pi} ds \, \int_0^{2\pi} d t \,\, \left[
 i {\rm Tr} \left\{ \hat P_- [ \partial_s  \hat P_-, \partial_t \hat P_- ] \right\}
+ i {\rm Tr} \left\{ \hat P_+ [ \partial_s  \hat P_+, \partial_t \hat P_+ ] \right\} \right]
= 2 \pi \times \text{integer} . 
\end{equation}
If (and only if) the U(1) bundle is trivial,  
these integrals of the curvature vanishes identically. 
The integrability condition eq.~(\ref{eq:integrability-condition}) asserts that it is indeed the case and 
the  fermion measure is then smooth.  
The global integrability condition discussed in the next subsection, on the other hand, asserts that
 the holonomy of the U(1) bundle 
is reproduced by the "Wilson line" of the connection.

\subsection{Proof of the reconstruction theorem}

\subsubsection{Global integrability condition}

As a first step to prove the reconstruction theorem, we formulate the so-called global 
integrability condition \cite{Luscher:1999un}. 

Let us assume that currents $j^a_\mu(x)$$(a=1,2,3)$ and $j_\mu(x)$ are local and 
satisfy all four properties required for the reconstruction theorem. 
We consider a definite topological sector $\mathfrak{U}^{(2)}[Q] \otimes \mathfrak{U}^{(1)}[0]$ and 
choose an arbitrary reference field $U^{(2)}_0\otimes U^{(1)}_0$ 
in this sector. Any other field 
$U^{(2)}\otimes U^{(1)}$ in the same sector can then be reached through a smooth curve 
$U_t$ such that $U_1=U^{(2)}\otimes U^{(1)}$. 
Then the basis vectors of the fermion fields at the point $U^{(2)}\otimes U^{(1)}$ 
may be chosen as follows \cite{Luscher:1999un}:
\begin{eqnarray}
\label{eq:chiral-basis-choice-v}
v_j(x) &=&\left\{    \begin{array}{ll}
   Q_{1-}  v^0_{1} \, W^{-1} &\quad \text{if $j=1$} , \\
   Q_{1-}  v^0_{j}                &\quad \text{otherwise} , 
   \end{array} \right. \\
u_j(x) &=&Q_{1+}  u^0_{j} , 
\end{eqnarray}
where $W$ is defined by\begin{equation}
W \equiv {\rm exp}\left\{ i \int_0^1 dt  \, \mathfrak{L}_\eta \right\}, 
   \qquad
\eta_\mu(x) = i \partial_t U_t(x,\mu) \, U_t(x,\mu)^{-1} , 
\end{equation}
$Q_{\pm t}$ is defined by the evolution operator of the projector
$P_{t \pm} = \left. \hat P_{\pm} \right\vert_{U=U_t}$ satisfying 
\begin{equation}
\label{eq:evolution-operators}
\partial_t Q_{t \pm} = \left[ \partial_t P_{t \pm} , P_{t \pm} \right] Q_{t \pm} ,   \quad Q_{0 \pm} = 1 , 
\end{equation}
and 
$u_j^0, v_{j}^0$ are the basis vectors for  the reference link field at $t=0$, 
$U^{(2)}_0 \otimes U^{(1)}_0$.  
The basis is path-dependent and, in general, the fermion measure defined with this basis
is also path-dependent.  In fact, any two curves $U_t$ and $\tilde U_t$ 
define two different sets of the basis vectors, $(v_j, u_j)$ and $(\tilde v_j, \tilde u_j)$, and 
the unitary transformation relating  them does not necessarily has determinant $1$. 
The fermion measure defined with the basis vectors is smooth if (and only if) it holds ture 
for any closed curve $U_t$ ($t \in [0,1]; U_1=U_0$) 
in the space $\mathfrak{U}^{(2)}[Q] \otimes \mathfrak{U}^{(1)}[0]$  that
\begin{equation}
\label{eq:global-integrability-condition}
W = {\rm det}(1-P_{0 -} + P_{0 -} Q_{1 -}) {\rm det}(1-P_{0 +} + P_{0 +} Q_{1 +}). 
\end{equation}
This condition is referred as global integrability condition.  The reconstruction theorem follows from 
the global integrability condition.

If a given closed curve is contractible,  the global integrability condition reduces to eq.~(\ref{eq:integrability-condition}), the local version of the integrability condition.  
Then, what is actually required by the global integrability condition is  that 
eq.~(\ref{eq:global-integrability-condition}) holds true
for any non-contractible loops in the space $\mathfrak{U}^{(2)}[Q] \otimes \mathfrak{U}^{(1)}[0]$. 
Moreover, with the smooth deformation of a given non-contractible loop, the global integrability condition holds true. In particular, the base point (the point at $t=0, 1$) of a non-contractible loop may be chosen  arbitrarily in the given 
topological sector $\mathfrak{U}^{(2)}[Q] \otimes \mathfrak{U}^{(1)}[0]$. 
Then, one may choose
$U_0=U^{(2)} \otimes 1$ 
with 
a certain  SU(2) link field in $\mathfrak{U}^{(2)}[Q] $ as the base point of non-contractible loops. 

\subsubsection{Non-contractible loops in the space of SU(2)$\times$U(1) gauge fields}

Since $\mathfrak{U}^{(2)}[Q] \otimes \mathfrak{U}^{(1)}[0]$ is a direct product space, 
any non-contractible loop in $\mathfrak{U}^{(2)}[Q] \otimes \mathfrak{U}^{(1)}[0]$ may be deformed 
to the product of the loops in $\mathfrak{U}^{(2)}[Q]$ and 
$\mathfrak{U}^{(1)}[0]$, respectively. Namely, one may assume that a non-contractible loop in $\mathfrak{U}^{(2)}[Q] \otimes \mathfrak{U}^{(1)}[0]$ has the following form (without loss of 
generality):
\begin{equation}
U_t = \left\{ 
\begin{array}{ll}
U^{(2)}_t \otimes 1 & ( 0 \le t \le 1 ; U^{(2)}_1=U^{(2)}_0 = U^{(2)}) , \\
U^{(2)} \otimes U^{(1)}_t &  ( 1 \le t \le 2 ; U^{(1)}_1=U^{(1)}_2 = 1 ), 
\end{array}
\right. 
\end{equation}
with a certain SU(2) link field $U^{(2)}$ in $\mathfrak{U}^{(2)}[Q] $. 
Then, in order to prove the global integrability condition,  one may consider separately the following two cases,  (1) {\it non-contractible loops in $\mathfrak{U}^{(2)}[Q] $ with the trivial 
U(1) link field as a background} and (2) {\it  non-contactible loops in $\mathfrak{U}^{(1)}[0]$ with 
an arbitrarily chosen SU(2) link field in $\mathfrak{U}^{(2)}[Q] $ as a background. }

%

In order to identify non-contractible loops in the topological sectors $\mathfrak{U}^{(2)}[Q] \otimes \mathfrak{U}^{(1)}[0]$, one needs to  clarify the topological structure of the space of the admissible 
SU(2)$\times$U(1)  gauge fields. 

As to the admissible U(1) gauge fields, 
it has been shown in \cite{Luscher:1998du} that 
the topological structure of $\mathfrak{U}^{(1)}[m]$ is a $(4+L^4-1)$-dimensional torus 
times a contractible space. 
Any admissible U(1) gauge field in a given topological sector $\mathfrak{U}^{(1)}[m]$ 
can be expressed as 
\begin{equation}
\label{eq:U-tilde}
U^{(1)}(x,\mu)=\tilde U^{(1)}(x,\mu) \, V_{[m]}(x,\mu) , 
\end{equation}
where 
\begin{eqnarray}
V_{[m]}(x,\mu) 
&=&{\text{e}}^{-\frac{2\pi i}{L^2}\left[
L \delta_{ \tilde x_\mu,L-1} \sum_{\nu > \mu} m_{\mu\nu}
 \tilde x_\nu +\sum_{\nu < \mu} m_{\mu\nu} \tilde x_\nu
\right]} .  
\end{eqnarray}
Here the abbreviation $\tilde x_\mu = x_\mu$ mod $L$ has been used. 
$V_{[m]}(x,\mu)$ has the constant field tensor equal to $2\pi m_{\mu\nu}/L^2 (< \epsilon_1)$
and may be regarded as a reference field of $\mathfrak{U}^{(1)}[m]$. 
Then $\tilde U^{(1)}(x,\mu)$ 
stands for the dynamical degrees of freedom  in the given topological sector. It 
can be parametrized with the three degrees of freedom:  
\begin{equation}
\label{eq:U-in-hodge-decomp}
\tilde U^{(1)}(x,\mu) = \Lambda(x)  \, {\rm e}^{i A^T_\mu(x)} \,  U_{[w]}(x,\mu) \, \Lambda(x+\hat\mu)^{-1}, 
\end{equation}
where 
$A^T_\mu(x)$ is the transverse vector potential  satisfying 
\begin{eqnarray}
&&\partial_\mu^\ast  A^T_\mu(x) = 0, \qquad
     \sum_{x\in \Gamma} A^T_\mu(x) = 0, \\
&& 
\partial_\mu A^T_\nu(x)-\partial_\nu A^T_\mu(x) + 2\pi m_{\mu\nu}/L^2 
= F_{\mu\nu}(x) . 
\end{eqnarray}
$U_{[w]}(x,\mu)$ represents the degrees of freedom of  the Wilson lines  defined by 
\begin{equation}
\label{eq:link-field-wilson-lines}
U_{[w]}(x,\mu) = \left\{ 
\begin{array}{cl}
w_\mu=\prod_{s=0}^{L-1} \{ \tilde U^{(1)}(0+s\hat\mu,\mu) \, {\rm e}^{-i A_\mu^T(0+s\hat\mu)} \} 
 &\quad  \text{if $x_\mu = L-1$},  \\
1           &\quad \text{otherwise,  }  
\end{array} \right.
\end{equation}
and $\Lambda(x)$ is the gauge function statisfying $\Lambda(0)=1$. 
By this parametrization, one can see that   
the space of the vector potentials $A^T_\mu(x)$, denoted by $\mathfrak{A}$, is contactible, 
while the space of the gauge functions $\Lambda(x)$, denoted by $\mathfrak{G}_0$, 
is $(L^4-1)$-dimensional torus. 
Therefore, the topological structure of $\mathfrak{U}^{(1)}[m]$ is a $(4+L^4-1)$-dimensional torus 
times a contractible space:
\begin{equation}
\mathfrak{U}^{(1)}[m] \simeq U(1)^4 \times \mathfrak{G}_0 \times \mathfrak{A} .  
\end{equation}
Then, one can see that there exist two kinds of non-contractible loops $( 0  \le t \le 1 )$ in $\mathfrak{U}^{(1)}[m]$. 
The first one is the gauge loops given by 
\begin{equation}
U_t^{(1)}(x,\mu) = \Lambda_t(x) V_{[m]}(x,\mu) \Lambda_t(x+\hat\mu)^{-1} ,  \quad 
\Lambda_t(x)= \exp\{ i 2 \pi t \delta_{\tilde x  \tilde y} \} . 
\end{equation}
The second one is the non-gauge loops given by
\begin{equation}
U_t^{(1)}(x,\mu) = V_{[m]}(x,\mu) \exp\{ i 2 \pi t \delta_{\mu\nu} \delta_{\tilde x  0} \} . 
\end{equation}

On the other hand, the topological structure of the space of the admissible SU(2) gauge fields, 
$\mathfrak{U}^{(2)}[Q]$, is not known so far \cite{Adams:2002ms}.  But, 
as long as one considers only the topological sectors with vanishing 
U(1) magnetic flux, $\mathfrak{U}^{(2)}[Q] \otimes \mathfrak{U}^{(1)}[0]$, 
one can establish the global integrability condition without the knowledge, by virtue of the pseudo 
reality of SU(2). 

\subsubsection{SU(2) loops  \ -- use of the pseudo reality of SU(2) --}

We first consider the case
(1) {\it non-contractible loops in $\mathfrak{U}^{(2)}[Q] $ with the trivial 
U(1) link field as a background}. 
When $U^{(1)}(x,\mu) = 1$, 
only the SU(2) gauge field couples to the left-handed fermion $\psi_-(x)$, which now consists of 
four degenerate SU(2) doublets. 
By noting the pseudo reality of SU(2), 
and the charge- and $\gamma_5$-conjugation properties of the lattice Dirac operator, 
one may choose the basis vectors of the left-handed fermion 
$\psi_-(x)$$={}^t\big($$q_-^1(x)$, $q_-^2(x)$, $q_-^3(x)$, $l_-(x)$$\big)$ 
for any given SU(2) gauge field $U^{(2)}(x,\mu) \in \mathfrak{U}^{(2)}[Q] $ as given 
by eqs.~(\ref{eq:four-doublets-basis-q1}), (\ref{eq:four-doublets-basis-q2}), 
(\ref{eq:four-doublets-basis-q3}) and (\ref{eq:four-doublets-basis-l}). 
With this choice of the basis,  one can infer that the measure term vanishes identically. 

For any closed curve in the space $\mathfrak{U}^{(2)}[Q] $, 
$U^{(2)}_t(x,\mu)$ ($t \in [0,1]$), 
one then has 
\begin{equation}
W = 1 . 
\end{equation}
On the other hand, from the hermiticity of $P_{t -}$, the unitarity of $Q_{t -}$ and the 
charge conjugation properties of $P_{t -}$ and $Q_{t -}$, it follows that
\begin{eqnarray}
&& P_{t -}=(\gamma_5  C^{-1}\otimes  i\sigma_2)  
  \{ P_{t -}\}^T  ( C \gamma_5 \otimes (i\sigma_2)^{-1} ) , \\
&& Q_{t -}  =(\gamma_5  C^{-1}\otimes  i\sigma_2)  
  \{ {Q_{t -}}^{-1}\}^T ( C \gamma_5 \otimes (i\sigma_2)^{-1} ). 
\end{eqnarray}
Then one can infer that
\begin{equation}
{\rm det}(1-P_{0 -} + P_{0 -} Q_{1 -}) = {\rm det}\big(1-P_{0 -} + P_{0 -} \{ Q_{1 -}\}^{-1} \big), 
\end{equation}
or 
\begin{equation}
\label{eq:twist-su2}
{\rm det}(1-P_{0 -} + P_{0 -} Q_{1 -}) = \pm 1. 
\end{equation}
Since $\psi_-(x)$ consists of 
four degenerate SU(2) doublets, $P_{t -}$ and $Q_{t -}$ factorize as 
\begin{equation}
P_{t -} = \prod_{i=1}^4 \otimes \, P^{(i)}_{t -}, \quad
Q_{t -}= \prod_{i=1}^4 \otimes \, Q^{(i)}_{t -}, 
\end{equation}
where $P^{(i)}_{t -}$ and $Q^{(i)}_{t -}$ $(i=1,2,3,4)$ are the projection- and the evolution-operators for the $i$-th component SU(2) doublet, respectively, and each set of the 
operators $P^{(i)}_{t -}$ and $Q^{(i)}_{t -}$
satisfies the same identity as eq.~(\ref{eq:twist-su2}).  Therefore one obtains\footnote{The right-handed fermion does not contribute 
the integrability condition in this case: ${\rm det}(1-P_{0 +} + P_{0 +} Q_{1 +})=1$.}
\begin{eqnarray}
{\rm det}(1-P_{0 -} + P_{0 -} Q_{1 -}) 
&=& \prod_{i=1}^4 {\rm det}\big(1-P^{(i)}_{0 -} + P^{(i)}_{0 -} Q^{(i)}_{1 -}\big)
\nonumber\\
&=& \left[ {\rm det}\big( 1-P^{(1)}_{0 -} + P^{(1)}_{0 -} Q^{(1)}_{1 -}\big) \right]^4 
\nonumber\\
&=& 1.  
\end{eqnarray}
Thus the global integrability condition holds true for any closed curves in $\mathfrak{U}^{(2)}[Q]$
with the trivial U(1) link field as a background. 

The measure of the chiral fermion $\psi_-(x)$ can be defined globally within $\mathfrak{U}^{(2)}[Q]$ 
and the lattice counterpart of the SU(2) global 
anomaly \cite{Neuberger:1998rn,Bar:2002sa,Bar:2000qa} is absent in this case.

\subsubsection{U(1) loops with SU(2) background}

We next consider the case (2) {\it  non-contactible loops in $\mathfrak{U}^{(1)}[0]$ with 
an arbitrarily chosen SU(2) link field in $\mathfrak{U}^{(2)}[Q] $ as a background. }

For gauge loops, one has 
\begin{equation}
\left. \mathfrak{L}_\eta = {\rm tr}\{ \gamma_5(1-aD)(y,y)\} \right\vert_{U^{(2)}\otimes U^{(1)}_t},  
\qquad
\eta_\mu(x) = -i \, U_t(x,\mu)^{-1} \partial_t U_t(x,\mu) = -\partial  \delta_{\tilde x  \tilde y},  
\end{equation}
where the SU(2) gauge field $U^{(2)}(x,\mu)$ is  chosen arbitrarily in $\mathfrak{U}^{(2)}[Q]$ 
and is fixed as a background.
Then the l.h.s.  is evaluated as  
\begin{equation} 
W = \exp\{ i 2 \pi[{\rm tr}\{ Y_- \gamma_5 (1-D_L)(y,y) \} 
- {\rm tr}\{ Y_+ \gamma_5 (1-D_L)(y,y) \}] \vert_{t=0}\} . 
\end{equation}
On the other hand, the factors in the r.h.s. are evaluated as 
\begin{eqnarray}
 \det\{1-P_{0\pm} + P_{0\pm} Q_{1\pm} \} 
&&=\lim_{n \rightarrow \infty} 
\det \left\{ 1-P_{0\pm} +(P_{0\pm} \Lambda_{\Delta t}^{-1} P_{0\pm} )^n \right\}  
\nonumber\\
&&=\exp\left\{ - i 2 \pi {\rm Tr}[\omega Y_\pm P_{0\pm} ]  \right\}, 
\end{eqnarray}
where $\Delta t =2\pi/n$ and  $\omega(x) = \delta_{\tilde x \tilde y}$ and therefore
\begin{eqnarray}
&& \det\{1-P_{0-} + P_{0-} Q_{1-} \} \det\{1-P_{0+} + P_{0+} Q_{1+} \}  \nonumber\\
&& =  \exp\left\{ - i 2 \pi {\rm Tr}[\omega Y_-P_{0-} ]  \right\} 
           \exp\left\{ - i 2 \pi {\rm Tr}[\omega Y_+ P_{0+} ]  \right\} 
      = W. 
\end{eqnarray}
Thus 
the global integrability condition holds ture for the gauge loops. 
%

For non-gauge loops, one has
\begin{equation}
\left. \mathfrak{L}_\eta = 2\pi j_\nu(0) \right\vert_{U^{(2)}\otimes U^{(1)}_t}, \qquad 
 \eta_{\mu}(x)_{(\nu)}  = -i \, U_{[w]}(x,\mu)^{-1} \partial_{t_\nu} U_{[w]}(x,\mu) 
= 2\pi \delta_{\mu\nu} \delta_{\tilde x 0}, 
\end{equation}
where again the SU(2) gauge field $U^{(2)}(x,\mu)$ is  chosen arbitrarily in $\mathfrak{U}^{(2)}[Q]$ 
and is fixed as a background.
Noting the charge conjugation properties of the U(1) measure term current under 
the transformation, $U_{[w]} \rightarrow U_{[w]}^\ast$, 
$U^{(2)} \rightarrow U^{(2)\ast} = (i\sigma_2) \, U^{(2)} \, (i\sigma_2)^{-1}$: 
\begin{equation}
j_\mu(x)\vert_{{U^{(1)}}^\ast, \, U^{(2)}} = + j_\mu(x)\vert_{U^{(1)}, \, U^{(2)}}, 
\end{equation}
the l.h.s. can be evaluated as 
\begin{equation}
W= \exp\left\{ i \int_0^{2\pi} dt j_\nu(0)  \right\}
= \exp\left\{ i \int_0^{\pi} dt j_\nu(0) - i \int_0^{-\pi} dt  j_\nu(0) \right\}
      = 1
\end{equation}
On the other hand, the r.h.s. can be evaluated as 
$(n=2r)$
\begin{eqnarray*}
&& \det\{1-P_{0\pm} + P_{0\pm} Q_{1\pm} \}  \\
&&=\lim_{n \rightarrow \infty} 
\det \left\{ 1-P_{0\pm} +
P_{0\pm}({\cal C}_\pm)^{-1}P_{t_1\pm} {\cal C}_\pm \cdots 
({\cal C}_\pm)^{-1}P_{t_r \pm} {\cal C}_\pm \right. \\
&& \left. \qquad \qquad \qquad \times P_{t_{r-1} \pm} P_{t_{r-2}\pm} \cdots P_{t_{1}\pm} P_{t_{0}\pm}  \right\} \\
&&=  \det \left\{1-P_{0\pm} +P_{0\pm} ({\cal C}_\pm)^{-1} P_{0\pm} \right\}
                         \det \left\{1-P_{\pi\pm} +P_{\pi\pm} ({\cal C}_\pm)^{-1} P_{\pi\pm} \right\} 
\end{eqnarray*}
where ${\cal C}_+= \left( \gamma_5 C^{-1} \right)$ and 
${\cal C}_-=\left( \gamma_5 C^{-1} \otimes  i \sigma_2 \right)$.  
Each factor in the final expression is $\pm 1$ because $\{{\cal C}_\pm\}^2=1$. 
The total expression is unity because, for the case of the right-handed factor, 
all SU(2) singlets have even hyper-charges and, 
for the left-handed factor,  
all four SU(2) doublets have odd hyper-charges.  
Thus the global integrability condition holds ture for the non-gauge loops, too. 

This completes the proof of the global integrability condition, and therefore,  the reconstruction theorem. 

\section{An explicit construction of the mesure term} 
\label{sec:measure-term-in-V}

In this section, we explicitly construct 
the local 
currents $j^a_\mu(x)$$(a=1,2,3)$ and $j_\mu(x)$ 
which satisfy all the required properties for the reconstruction theorem
in the topological sectors $\mathfrak{U}^{(2)}[Q] \otimes \mathfrak{U}^{(1)}[0]$
with vanishing magnetic fluxes $m_{\mu\nu}=0$. 
We follow the approach in our previous work for the U(1) case \cite{Kadoh:2007},  
extending the construction there to the case of the SU(2)$\times$U(1) chiral gauge theory. 



%

\subsection{Parametrization of U(1) link fields and their variations in finite volume}

We fisrt discuss the parametrization of the link fields in finite volume and their variations. 
We adopt the parametrization of the U(1) link fields given by eqs.~(\ref{eq:U-tilde}) and 
(\ref{eq:U-in-hodge-decomp}).  
It is unique and the each factors, 
$A_\mu^T(x)$, $U_{[w]}(x,\mu)$ and $\Lambda(x)$, may be regarded as 
the smooth functionals of the original link field $U^{(1)}(x,\mu)$.  

Accordingly, 
the variation of the U(1) link field, 
\begin{equation}
\delta_\eta U^{(1)}(x,\mu) = i \, \eta_\mu(x) \, U^{(1)}(x,\mu), 
\end{equation}
may be decomposed as follows:
\begin{equation}
\eta_\mu(x) = \eta^T_\mu(x) + \eta_{\mu [w]}(x) + \eta_\mu^\Lambda(x) .   
\end{equation}
$\eta^T_\mu(x)$ is the transverse part of $\eta_\mu(x)$ defined by
\begin{equation}
\partial_\mu^\ast \eta^T_\mu(x) = 0, \qquad \sum_{x \in \Gamma} \eta^T_\mu(x) =0, 
\end{equation}
which may be given explicitly as
\begin{equation}
\eta^T_\mu(x)  = \sum_{y\in\Gamma} G_L(x-y) 
\partial_\lambda^\ast (\partial_\lambda \eta_\mu(x) - \partial_\mu \eta_\lambda(x)).  
\end{equation}
$\eta_{\mu [w]}(x)$ is the variation along the Wilson lines defined by 
\begin{equation}
\label{eq:variation-parameter-wilson-lines}
\eta_{\mu [w]}(x) =\sum_{\nu} \eta_{(\nu)} \, \delta_{\mu\nu} \, \delta_{x_\nu,L-1} , \qquad
 \eta_{(\nu)} = L^{-3} \sum_{y\in \Gamma} \eta_\nu(y) .  
\end{equation}
$\eta_\mu^\Lambda(x)$ is the variation of the gauge degrees of freedom in the form,
\begin{equation}
\eta_\mu^\Lambda(x) = - \partial_\mu \omega_\eta(x) ,  \qquad \omega_\eta(0) = 0.
\end{equation}
This decomposition is also unique by the following reason:  for an arbitrary periodic vector field 
$\eta_\mu(x)$, the vector field defined by 
$a_\mu(x)=\eta_\mu(x)-\eta_\mu^T(x) - \eta_{ \mu [w]}(x)$ has the vanishing field tensor
$\partial_\mu a_\nu(x) - \partial_\nu a_\mu(x)=0$ and the vanishing wilson lines
$\sum_{s=0}^{L-1} a_\mu(x+s \hat \mu)=0$. 
Then, the sum $\omega_\eta(x)$ of the vector field
$a_\mu(x)$ along any lattice path from $x$ to the origin $x=0$ is independent of the
chosen path,  periodic in $x$ and $\omega_\eta(0)=0$.  It gives the gauge function which reproduces 
$a_\mu(x)$ in the pure gauge form, $a_\mu(x) = -\partial_\mu \omega_\eta(x)$.  This proves the uniqueness of the decomposition. 
The action of the differential operator $\delta_\eta$ to each factors, $A_\mu^T(x)$, 
$U_{[w]}(x,\mu)$ and $\Lambda(x)$, is then given as follows:
\begin{eqnarray}
&& \delta_\eta A_\mu^T(x) = \eta_\mu^T(x) , \\
&& \delta_\eta U_{[w]}(x,\mu) = i \, \eta_{\mu [w]}(x) \, U_{[w]}(x,\mu) , \\
&& \delta_\eta \Lambda(x)  
= i \, \omega_\eta(x) \,  \Lambda(x) . 
\end{eqnarray}

\subsection{A closed formula of the measure term in finite volume}

We now give an explicit formula of  the measure term 
for the admissible SU(2)$\otimes$U(1) gauge fields in the topological sectors 
$\mathfrak{U}^{(2)}[Q] \otimes \mathfrak{U}^{(1)}[0]$ with the vanishing magnetic 
fluxes $m_{\mu\nu}=0$.\footnote{A general strategy to construct the SU(2) part of the 
measure term was discussed 
in \cite{Kikukawa:2001jm}. We follow this strategy,  specifying explicitly the U(1) part of the measure 
term current $j_\mu(x)$ and the interpolation path in the U(1) direction. }  
For this purpose,  we introduce a vector potential defined by
\begin{eqnarray}
&& \tilde A_\mu^\prime(x) = A_\mu^T(x) - \frac{1}{i}\partial_\mu \Big[ \ln \Lambda(x) \Big] ; \qquad
\frac{1}{i} \ln \Lambda(x) \in (-\pi, \pi] ,  
\end{eqnarray}
and choose a  one-parameter family of the gauge fields as 
\begin{equation}
\label{eq:family-U-s}
U_s(x,\mu)= U^{(2)}(x,\mu) \otimes \left[ {\rm e}^{i s \tilde A_\mu^\prime(x)} \,  U_{[w]}(x,\mu) \right] , 
\qquad   0 \le s \le 1 .
\end{equation}
%
%
Then we consider the linear functional of the variational parameters
$\eta^{(2)}_\mu(x)$ and $\eta^{(1)}_\mu(x)$ given by
\begin{eqnarray}
\label{eq:measure-term-finite-volume}
\mathfrak{L}_\eta^\diamond &=&
   i \int_0^1 ds \, {\rm Tr} \left\{ \hat P_- [ \partial_s \hat P_-, \delta_\eta \hat P_- ] \right\} 
+ i \int_0^1 ds \, {\rm Tr} \left\{ \hat P_+ [ \partial_s \hat P_+, \delta_\eta \hat P_+ ] \right\} 
\nonumber\\
&& +
\, \delta_\eta \, 
\int_0^1 ds \, \sum_{x \in \Gamma_4} 
 \left\{   \tilde A_\mu^\prime(x)  \,   k_\mu(x) \right\}   
+ {\mathfrak{L}}_\eta \vert_{U=U^{(2)} \otimes U_{[w]}} ,
\end{eqnarray}
where $k_\mu(x)$ is the gauge-invariant local current
which satisfies 
$\partial_\mu^\ast  k^{}_\mu(x) = q^{(1)}_L(x)$ and 
transforms as an axial vector field under the lattice symmetries. 
The additional term 
${\mathfrak{L}}_\eta \vert_{U=U^{(2)} \otimes U_{[w]}}$ 
is the measure term at the gauge fields $U_0(x,\mu)=U^{(2)}(x,\mu) \otimes U_{[w]}(x,\mu)$,
which construction will be dicussed in the following section~\ref{subsec:measure-term-at-su2xu1_wilson_lines}. 

The currents $j^{a \diamond}_\mu(x)$$(a=1,2,3)$, $j_\mu^\diamond(x)$
defined by 
eq.~(\ref{eq:measure-term-finite-volume}), 
\begin{equation}
{\mathfrak{L}}_\eta^\diamond = 
\sum_{x\in \Gamma} \{ \eta^a_\mu(x) j^{a  \diamond}_\mu(x)+ \eta_\mu(x) j_\mu^\diamond(x) \}, 
\end{equation}
may be regarded as a functional of the link variable $U(x,\mu)$ through the dependences on
$U^{(2)}(x,\mu)$, 
$A_\mu^T(x)$, $\Lambda(x)$ ($\ln \Lambda(x)$) and $U_{[w]}(x,\mu)$. 
The action of the differential operator $\delta_\eta$ to 
the vector potential $\tilde A_\mu^\prime(x)$ is evaluated as
\begin{eqnarray}
\delta_\eta \tilde A_\mu^\prime(x)
&=& 
\delta_\eta  A_\mu^T(x) - \partial_\mu \left[\frac{1}{i} \{ \delta_\eta \Lambda(x) \}\, \Lambda(x)^{-1} \right] 
=\eta_\mu^T(x) - \partial_\mu \omega_\eta(x) \nonumber\\
&=&
\eta_\mu(x)-\eta_{\mu [w]}(x) , 
\end{eqnarray}
and the variation of $U_s(x,\mu)$ is given by 
\begin{eqnarray}
\delta_\eta U_s(x,\mu) &=&
i \eta^{(2)}_\mu (x) \, U^{(2)}(x,\mu) \otimes \left[ {\rm e}^{i s \tilde A_\mu^\prime(x)} \,  U_{[w]}(x,\mu) \right]
\nonumber\\
&+&
 U^{(2)}(x,\mu) \otimes
 i \left\{
 s(\eta^{(1)}_\mu(x) -\eta_{\mu [w]}(x)  ) + \eta_{\mu [w]}(x) \right\}
 \left[ {\rm e}^{i s \tilde A_\mu^\prime(x)} \,  U_{[w]}(x,\mu) \right] .  \nonumber\\
\end{eqnarray}

The linear functional $\mathfrak{L}_\eta^\diamond $
so obtained, however, does not respect the lattice symmetries. 
In order to make it
to transform as a pseudo scalar field under the lattice 
symmetries, we should average it over the lattice symmetries with the appropriate 
weights so as to project to the pseudo scalar component. 
Namely, we take the average as follows\footnote{In doing the average, one should note the fact that 
under the lattice symmetries the Wilson lines $U_{[w]}(x,\mu)$ 
are transformed  to other Wilson lines $U_{[w^\prime]}(x,\mu) $ {\em modulo gauge transformations}, 
$\{ U_{[w]}(x,\mu)\}^{R^{-1}}  = U_{[w^\prime]}(x,\mu) \Lambda(x) \Lambda(x+\hat \mu)^{-1}$. 
Accordingly, the variational parameter $\eta_{\mu [w]}(x)$ is transformed as 
$\{\eta_{\mu [w]}(x)\}^{R^{-1}} = \eta_{\mu [w^\prime]}(x) - \partial_\mu \omega(x)$ with a certain 
periodic gauge funciton $\omega(x)$. }:
\begin{equation}
\label{eq:average-L-diamond}
\bar{\mathfrak{L}}^\diamond_{\eta} = \frac{1}{2^4 4!} \sum_{R \in  O(4,\mathbb{Z})} \det R \, \, 
 \mathfrak{L}^\diamond_{\eta}\vert_{U\rightarrow\{U\}^{R^{-1}},  
         \eta_\mu\rightarrow\{\eta_\mu \}^{R^{-1}} } . 
\end{equation}

Our main result is then stated as follows:

\vspace{1em}
\noindent{\bf Lemma 2} \ \ {\sl The currents 
$j^{a \diamond}_\mu(x)$$(a=1,2,3)$, $j_\mu^\diamond(x)$ defined by 
eq.~(\ref{eq:measure-term-finite-volume}), 
\begin{equation*}
{\mathfrak{L}}_\eta^\diamond = 
\sum_{x\in \Gamma} \{ \eta^a_\mu(x) j^{a  \diamond}_\mu(x)+ \eta_\mu(x) j_\mu^\diamond(x) \}, 
\end{equation*}
fulfills all the properties required for the reconstruction theorem in the lattice Glashow-Weinberg-Salam model except the transformation property 
under the lattice symmetries. It may be corrected by invoking the average 
eq.~(\ref{eq:average-L-diamond}) over 
the lattice symmetries with the appropriate 
weights so as to project to the pseudo scalar component. 
}

\vspace{1em}
\noindent
The proof of this statement will be given in section~\ref{subsec:proof-lemma-4a}.  The locality property of the currents will be examined in section~\ref{sec:locality-of-currents}. 

\subsection{Measure term at $U^{(2)}(x,\mu) \otimes U_{[w]}(x,\mu)$}
\label{subsec:measure-term-at-su2xu1_wilson_lines}

The measure term at the gauge fields $U(x,\mu)=U^{(2)}(x,\mu) \otimes U_{[w]}(x,\mu)$
should consist of the two components:
\begin{equation}
\mathfrak{L}_\eta \vert_{U=U^{(2)} \otimes U_{[w]}}  = \left\{ 
\begin{array}{ll}
\mathfrak{L}_{\eta} \vert_{U=U^{(2)} \otimes U_{[w]};\eta=\eta_{[w]}} 
&  \text{ for } \ \eta_\mu(x)=\eta^{(1)}_\mu(x)=\eta_{\mu[w]}(x),  \\
\mathfrak{L}_\eta \vert_{U=U^{(2)} \otimes U_{[w]};\eta=\eta^{(2)}} 
&   \text{ for } \ \eta_\mu(x)=\eta^{(2)}_\mu(x).  \\
\end{array} \right.
\end{equation}


In order to construct the measure term at the gauge field 
$U(x,\mu)=U^{(2)}(x,\mu)\otimes U_{[w]}(x,\mu)$
with the variational parameters in the directions of the U(1) Wilson lines $\eta_{\mu [w]}(x)$, 
we first discuss a special property of 
the curvature terms associated with the U(1) Wilson lines $U_{[w]}(x,\mu)$, which turn out to be 
useful in the construction of 
a solution to the integrability condition eq.~(\ref{eq:integrability-condition}). 
Let us parametrize the Wilson lines $U_{[w]}(x,\mu) $ defined by eq.~(\ref{eq:link-field-wilson-lines}) as  
\begin{equation}
\label{eq:wilson-lines-parameter}
w_\nu = \exp( i t_\nu ),  \quad  t_\nu \in [0, 2\pi)  \quad (\nu =1,2,3,4) , 
\end{equation}
and the variational parameters in the directions of the Wilson lines as
\begin{equation}
\label{eq:wilson-lines-variation}
\lambda_{\mu(\nu)}(x) = \frac{1}{i} \, \partial_{t_\nu} U_{[w]}(x,\mu) \cdot U_{[w]}(x,\mu)^{-1}  
= \delta_{\mu\nu} \delta_{x_\nu,L-1}.    
\end{equation}
The curvature term for the Wilson lines reads
\begin{equation}
\left[
 i {\rm Tr} \big\{
\hat P_+ [ \partial_{t_\mu} \hat P_+ , \partial_{t_\nu} \hat P_+ ] \big\} 
+i {\rm Tr} \big\{
\hat P_- [ \partial_{t_\mu} \hat P_- , \partial_{t_\nu} \hat P_- ] \big\} 
\right]_{U=U^{(2)} \otimes U_{[w]} V_{[m]}}
\equiv  \mathfrak{C}_{\mu\nu}(t), 
\end{equation}
where $t= (t_1,t_2,t_3,t_4)$. 
Then the following lemma holds true:

\vspace{1.5em}
\noindent{\bf Lemma 3} \ \ 
{\sl 
In the topological sectors $\mathfrak{U}^{(2)}[Q] \otimes \mathfrak{U}^{(1)}[m]$
of the lattice Glashow-Weinberg-Salam model, 
the curvature term for the U(1) Wilson 
lines $\mathfrak{C}_{\mu\nu}(t)$, 
which possesses the properties 
\begin{equation}
\label{eq:properties-of-curvature-1}
\mathfrak{C}_{\mu\nu}(t)  = - \mathfrak{C}_{\nu\mu}(t), \qquad
\partial_\mu \mathfrak{C}_{\nu\lambda}(t) 
+\partial_\nu \mathfrak{C}_{\lambda\mu}(t) 
+\partial_\lambda \mathfrak{C}_{\mu\nu}(t) 
= 0, 
\end{equation}
satisfies the bound
\begin{equation}
\label{eq:properties-of-curvature-2}
\left\vert \mathfrak{C}_{\mu\nu} (t) \right\vert  \le  \kappa L^\sigma {\rm e}^{-L/\varrho}
\end{equation}
for certain positive constants $\kappa$ and $\sigma$, while $\varrho$ is the 
localization range of the lattice Dirac operator $D$.
For a sufficiently large volume $L^4$, 
it then follows that
\begin{equation}
\label{eq:properties-of-curvature-3}
\int_0^{2\pi} d t_\mu \, \int_0^{2\pi} d t_\nu \,\, \mathfrak{C}_{\mu\nu}(t)  =0,  
\end{equation}
and  there exists smooth periodic vector field $\mathfrak{W}_\mu(t)$ such that 
\begin{equation}
\mathfrak{C}_{\mu\nu}(t)= \partial_\mu \mathfrak{W}_\nu(t)
                                         - \partial_\nu \mathfrak{W}_\mu(t), \qquad
\left| \mathfrak{W}_\mu(t) \right| \le 3 \pi  \, \text{sup}_{t,\mu,\nu} \left| \mathfrak{C}_{\mu\nu}(t) \right| . 
\end{equation}
}

\vspace{1em}
\noindent
The proof of this lemma has been given for the U(1) case in \cite{Kadoh:2007}, which holds true also
for the SU(2)$\otimes$U(1) case here by  regarding the SU(2) gauge field in the background.
The proof   
is based on the fact that
in infinite-volume 
the periodic link field which represents the degrees of freedom of the Wilson lines
can be written in the pure-gauge form, 
\begin{equation}
\label{eq:wilon-line-in-pure-gauge}
U_{[w]}(x,\mu) = \Lambda_{[w]}(x) \Lambda_{[w]}(x+\hat \mu)^{-1} ,   \quad 
\Lambda_{[w]}(x)= \prod_\mu ( w_\mu )^{n_\mu} \quad \text{for  } x- n L \in \Gamma, 
\end{equation}
and therefore
the gauge-invariant function of the link field in infinite volume is actually independent of the 
degrees of freedom of the Wilson lines.  
In fact, 
noting the representation eq.~(\ref{eq:DL-in-D-infity}), 
one may rewrite the curvature term eq.~(\ref{eq:curvature-term})
into 
\begin{eqnarray}
\label{eq:curvature-infinite-finite}
\mathfrak{C}_{\eta\zeta}&=&
 i {\rm Tr} \left\{ Q_\Gamma \hat P_- [ \delta_\eta \hat P_-, \delta_\zeta \hat P_- ] \right\} 
+ i {\rm Tr} \left\{ Q_\Gamma \hat P_+ [ \delta_\eta \hat P_+, \delta_\zeta \hat P_- ] \right\} 
+ \mathfrak{R}_{\eta \zeta},  
\end{eqnarray}
where  
$Q_\Gamma$ here is the projector  acting on the fields in infinite volume as
\begin{equation}
Q_\Gamma \, \psi(x) = \left\{ 
\begin{array}{ll} \psi(x) & \text{if $x \in \Gamma, $} \\ 0 & \text{otherwise. } \end{array} \right.
\end{equation}
$\mathfrak{R}_{\eta \zeta}$  is the finite-volume correction to the curvature term given by
\begin{eqnarray}
&& \mathfrak{R}_{\eta \zeta} = 
i \sum_{s=\mp}\sum_{x\in\Gamma}  \sum_{y,z \in\mathbb{Z}^4} \sum_{n \in \mathbb{Z}^4, n \not = 0} 
{\rm tr}\left\{
P_s(x,y) \right. \nonumber\\
&&\qquad\qquad
\times \left. \left[
\delta_\eta P_s(y,z) \delta_\zeta P_s(z,x+Ln)- \delta_\zeta P_s(y,z) \delta_\eta P_s(z,x+Ln)
\right]\right\} , 
\end{eqnarray}
while $P_s(x,y) (s=\mp)$ are the kernels of the chiral projectors in infinite volume,
\begin{equation}
P_\mp(x,y) = \frac{1}{2}(1\mp\gamma_5)\delta_{xy} \pm  \frac{1}{2}\gamma_5 D(x,y).
\end{equation}
From eq.~(\ref{eq:locality-D}), one can infer  that 
\begin{eqnarray}
\label{eq:bound-R}
&& \left| \mathfrak{R}_{\eta \zeta} \right|  \le 
\kappa_1 L^{\nu_1} {\rm e}^{-L/\varrho} \| \eta \|_\infty \| \zeta \|_\infty
\end{eqnarray}
for some constants $\kappa_1 >0$ and $\nu_1 \ge 0$.  
We then recall the fact that there exists the measure term in infinite volume \cite{Kikukawa:2000kd}, 
$\mathfrak{K}_\eta = 
\sum_{x \in \Gamma} \left\{ \eta^a_\mu(x) j_\mu^{\star a}(x) + \eta_\mu(x) j_\mu^{\star}(x) \right\}$,  
which satisfies the integrability condition  
\begin{equation}
\label{eq:integrability-condition-infinite-volume}
 i {\rm Tr} \left\{ Q_\Gamma \hat P_- [ \delta_\eta \hat P_-, \delta_\zeta \hat P_- ] \right\} 
+ i {\rm Tr} \left\{ Q_\Gamma \hat P_+ [ \delta_\eta \hat P_+, \delta_\zeta \hat P_- ] \right\} 
=  \delta_\eta \mathfrak{K}_\zeta - \delta_\zeta \mathfrak{K}_\eta +  \mathfrak{K}_{[\eta,\zeta]}. 
\end{equation}
The currents  $j_\mu^{\star a}(x)$ and $j_\mu^\star(x)$ are defined for all admissible gauge fields in 
infinite volume and it is local and gauge-invariant under the U(1) gauge transformations.  ($j_\mu^{\star a}(x)$ and $j_\mu^\star(x)$ are gauge-covariant and gauge-invariant, respectively, under the SU(2) gauge transformation.)
Then, as discussed above,
the currents  
are actually independent of the 
Wilson lines and the curvature of $\mathfrak{K}_\eta$ 
evaluated  in the directions of the Wilson lines
vanishes identically. Namely, 
\begin{eqnarray}
&&
\left[ \delta_{\lambda_{(\mu)}}  \mathfrak{K}_{\lambda_{(\nu)}}
-\delta_{\lambda_{(\nu)}}  \mathfrak{K}_{\lambda_{(\mu)}} \right]_{U = U^{(2)}\otimes U_{[w]} V_{[m]} } 
\nonumber\\
&=& 
\left. 
 i {\rm Tr} \left\{ Q_\Gamma \hat P_- [ \delta_{\lambda_{(\mu)}} \hat P_-, \delta_{\lambda_{(\nu)}}
\hat P_- ] \right\} 
+ i {\rm Tr} \left\{ Q_\Gamma \hat P_+ [ \delta_{\lambda_{(\mu)}} \hat P_+, \delta_{\lambda_{(\nu)}}
\hat P_+ ] \right\} 
\right\vert_{U = U^{(2)}\otimes U_{[w]} V_{[m]} }  \nonumber\\
&=& \mathfrak{C}_{\mu\nu}(t) -\mathfrak{R}_{\lambda_{(\mu)} \lambda_{(\nu)}}
= 0 .
\end{eqnarray}
This fact immediately implies that
the curvature term  for the Wilson lines, $\mathfrak{C}_{\mu\nu}$,
itself satisfies the bound eq.~(\ref{eq:properties-of-curvature-2})
and because of this bound, the two-dimensional integration of the curvature, 
which should be a multiple of $2\pi$, 
must vanish identically for a sufficiently large $L$.
The existence of the smooth periodic vector field $\mathfrak{W}_\mu(t)$ then follows from 
the lemma 9.2 in \cite{Luscher:1998du}. 
%

%
%

By the above lemma, 
one can construct  a solution of the integrability condition 
at the gauge fields $U(x,\mu)=U^{(2)}(x,\mu) \otimes U_{[w]}(x,\mu)$,  
\begin{equation}
 \delta_{\lambda_{(\mu)}} {\mathfrak{W}}_\nu
-\delta_{\lambda_{(\nu)}} {\mathfrak{W}}_\mu
=  \left.   \mathfrak{C}_{\mu\nu}  \right\vert_{U = U^{(2)}\otimes U_{[w]}}, 
\end{equation}
from $\mathfrak{C}_{\mu\nu}$ directly. 
The solution may be given explicitly by the formulae, 
\begin{eqnarray}
\mathfrak{W}_4
&=& \frac{1}{2\pi} \int_0^{2\pi} d r_4  \int_0 ^{(t_1,t_2,t_3)}  
\{ dr_1 \mathfrak{C}_{14} +dr_2 \mathfrak{C}_{24} +dr_3 \mathfrak{C}_{34}  \} ,  \nonumber\\
\mathfrak{W}_3
&=& 
\int_0^{t_4} dr_4  \mathfrak{C}_{43}  - \frac{t_4}{2\pi} \int_0^{2\pi} dr_4  \mathfrak{C}_{43}
+
\left[ \frac{1}{2\pi} \int_0^{2\pi} d r_3  \int_0 ^{(t_1,t_2)}  
\{ dr_1 \mathfrak{C}_{13} +dr_2 \mathfrak{C}_{23} \} \right]_{t_4=0} , 
 \nonumber\\
\mathfrak{W}_2
&=& 
 \int_0^{t_4} dr_4  \mathfrak{C}_{42}  - \frac{t_4}{2\pi} \int_0^{2\pi} dr_4  \mathfrak{C}_{42} \nonumber\\
 &&
+\left[ \int_0^{t_3} dr_3  \mathfrak{C}_{32}  
         - \frac{t_3}{2\pi} \int_0^{2\pi} dr_3  \mathfrak{C}_{32} \right]_{t_4=0} 
+\left[ \frac{1}{2\pi} \int_0^{2\pi} d r_2  \int_0 ^{(t_1)}  \{ dr_1 \mathfrak{C}_{12} \} \right]_{t_4=t_3=0} , 
\nonumber\\
\mathfrak{W}_1
&=& \int_0^{t_4} dr_4  \mathfrak{C}_{41}  - \frac{t_4}{2\pi} \int_0^{2\pi} dr_4  \mathfrak{C}_{41} \nonumber\\
&&+\left[\int_0^{t_3} dr_3  \mathfrak{C}_{31}  
             - \frac{t_3}{2\pi} \int_0^{2\pi} dr_3  \mathfrak{C}_{31}\right]_{t_4=0} 
+\left[\int_0^{t_2} dr_2  \mathfrak{C}_{21}  
        - \frac{t_2}{2\pi} \int_0^{2\pi} dr_2  \mathfrak{C}_{21}\right]_{t_4=t_3=0} . \nonumber\\
        \label{eq:measure-term-Wilson-lines}
\end{eqnarray}
It follows from the properties of $\mathfrak{C}_{\mu\nu}$
that this solution is periodic and smooth
with respect to the Wilson lines $U_{[w]} $ and 
satisfies the bound
\begin{equation}
\label{eq:measure-term-Wilson-lines-bound}
\left\vert \, {\mathfrak{W}}_\mu  \, \right\vert 
 \le  \kappa_2 L^{\nu_2} {\rm e}^{-L/\varrho},  
\end{equation}
for certain positive constants $\kappa_2$ and $\nu_2$. 
It also follows that this solution is gauge invariant. 
Then the linear functional 
of the variational parameters $\eta_{\mu[w]}(x)$, $ \sum_\nu \eta_{(\nu)} {\mathfrak{W}}_\nu$, 
provides the measure term at the gauge field 
$U(x,\mu)=U^{(2)}(x,\mu)\otimes U_{[w]}(x,\mu)V_{[m]}(x,\mu)$
with the variational parameters in the directions of the U(1) Wilson lines $\eta_{\mu [w]}(x)$:
\begin{equation}
{\mathfrak{L}}_\eta \vert_{U=U^{(2)} \otimes U_{[w]}; \eta=\eta_{[w]}}
= \sum_\nu \eta_{(\nu)} {\mathfrak{W}}_\nu . 
\end{equation}


On the other hand, 
the measure term at the gauge field 
$U(x,\mu)=U^{(2)}(x,\mu)\otimes U_{[w]}(x,\mu)$
 with the variational parameters in the directions of the SU(2) gauge fields
may be given by the following formulae:
\begin{eqnarray}
\label{eq:measure-term-Wilson-lines-su2}
\mathfrak{L}_\eta \vert_{U=U^{(2)} \otimes U_{[w]}; \eta=\eta^{(2)}} 
&=& \int_0^{t_1}  d r_1 \,  \mathfrak{C}_{1 \eta}(r_1,0,0,0)
 +\int_0^{t_2}  d r_2 \,  \mathfrak{C}_{2 \eta}(t_1,r_2,0,0)  \nonumber\\
&+& 
  \int_0^{t_3}  d r_3 \,  \mathfrak{C}_{3 \eta}(t_1,t_2,r_3,0) 
+\int_0^{t_4}  d r_4 \,  \mathfrak{C}_{4 \eta}(t_1,t_2,t_3,r_4) 
- \delta_\eta  \phi_{[w]} , \nonumber\\
\end{eqnarray}
and
\begin{eqnarray}
\phi_{[w]} &=& 
  \int_0 ^{(t_1)} dr_1 \, \mathfrak{W}_1(r_1,0,0,0)
+\int_0 ^{(t_2)} dr_2 \, \mathfrak{W}_2(t_1,r_2,0,0)  \nonumber\\
&&
+\int_0 ^{(t_3)} dr_3 \, \mathfrak{W}_3(t_1,t_2,r_3,0)
+\int_0 ^{(t_4)} dr_4 \, \mathfrak{W}_4(t_1,t_2,t_3,r_4). 
\end{eqnarray}
It is not difficult to show that the measure term so constructed indeed satisfies the integrability 
condition for all possible directions of the variational parameters. 
It also follows from the properties of $\mathfrak{C}_{\mu\nu}$ and $\mathfrak{C}_{\mu\eta^{(2)}}$
that 
the measure term current is  gauge covariant and  
is  periodic and smooth with respect to the Wilson lines $U_{[w]}(x,\mu)$. 
To see the latter property explicitly, 
one may rewrite the above formula as follows:
\begin{eqnarray*}
&& \mathfrak{L}_\eta \vert_{U=U^{(2)} \otimes U_{[w]}; \eta=\eta^{(2)}} \nonumber\\
&&= \int_0^{t_1}  d r_1 \, \mathfrak{C}_{1 \eta}(r_1,0,0,0)  \nonumber\\
&&+ \int_0^{t_2}  d r_2 \, \mathfrak{C}_{2 \eta}(t_1,r_2,0,0)  
-\frac{t_2}{2\pi} \int_0^{2\pi} d r_2   \mathfrak{C}_{2 \eta}(t_1,r_2,0,0) 
+\frac{t_2}{2\pi} \int_0^{2\pi} d r_2   \mathfrak{C}_{2 \eta}(0,r_2,0,0)
\nonumber\\
&&+ \int_0^{t_3}  d r_3 \, \mathfrak{C}_{3 \eta}(t_1,t_2,r_3,0) 
-  \frac{t_3}{2\pi} \int_0^{2\pi} d r_3 \, \mathfrak{C}_{3 \eta}(t_1,t_2,r_3,0)  
+  \frac{t_3}{2\pi} \int_0^{2\pi} d r_3 \, \mathfrak{C}_{3 \eta}(0,0,r_3,0)  
\nonumber\\
&&+ \int_0^{t_4}  d r_4 \, \mathfrak{C}_{4 \eta}(t_1,t_2,t_3,r_4) 
- \frac{t_4}{2\pi} \int_0^{2\pi} d r_4  \, \mathfrak{C}_{4 \eta}(t_1,t_2,t_3,r_4)
+ \frac{t_4}{2\pi} \int_0^{2\pi} d r_4  \, \mathfrak{C}_{4 \eta}(0,0,0,r_4). 
\end{eqnarray*}
Then, from  the pseudo reality of SU(2) and the charge conjugation property of 
$\mathfrak{C}_{\nu \eta}$ $(\nu=1,2,3,4)$,
\begin{equation}
\mathfrak{C}_{\nu \eta} \vert_{U=U^{(2)}\otimes \{U_t^{(1)}\}^\ast} 
= \mathfrak{C}_{\nu \eta} \vert_{U=U^{(2)}\otimes {U_t^{(1)}}}, 
\end{equation}
one can infer that 
\begin{eqnarray}
\int_0^{2\pi} d r_\nu  \, \mathfrak{C}_{\nu \eta}(0,\cdots,r_\nu, \cdots,0)
&=& \int_0^\pi d r_\nu \, \left[ \mathfrak{C}_{\nu \eta}(0,\cdots,r_\nu, \cdots,0) 
                                              -\mathfrak{C}_{\nu \eta}(0,\cdots,-r_\nu, \cdots,0) \right] \nonumber\\
&=& 0  \qquad (\nu=1,2,3,4).  
\end{eqnarray}
With these identities, one can easily verify that the measure term 
is periodic and smooth with respect to the Wilson lines $U_{[w]}(x,\mu)$. 

Finally, we note that the measure term ${\mathfrak{L}}_\eta \vert_{U=U^{(2)} \otimes U_{[w]}}$ 
so constructed satisfies the bound 
\begin{equation}
\label{eq:measure-term-Wilson-lines-bound2}
\left\vert \, \mathfrak{L}_{\eta} \vert_{U=U^{(2)} \otimes U_{[w]}}  \, \right\vert 
 \le  \kappa^\prime L^{\sigma^\prime} {\rm e}^{-L/\varrho} \, \|\eta \|_\infty
\end{equation}
for certain positive constants $\kappa^\prime$ and $\sigma^\prime$. 
For $\eta_\mu(x) = \eta_{\mu [w]}(x)$,  it immediately follows from eq.~(\ref{eq:measure-term-Wilson-lines-bound}). For $\eta_\mu(x) = \eta^{(2)}_\mu(x)$,  as one can see from the argument given in the proof of the lemma 4.b and 
eqs.~(\ref{eq:wilon-line-in-pure-gauge}) and (\ref{eq:integrability-condition-infinite-volume}), 
the curvature term $\mathfrak{C}_{\eta\zeta}-\mathfrak{R}_{\eta\zeta}$ does not  actually depend on the U(1) Wilson lines. 
Then, one may write 
\begin{eqnarray*}
\mathfrak{L}_\eta \vert_{U=U^{(2)} \otimes U_{[w]}; \eta=\eta^{(2)}} 
&&= \int_0^{t_1}  d r_1 \, \mathfrak{R}_{1 \eta}(r_1,0,0,0)  \nonumber\\
&&+ \int_0^{t_2}  d r_2 \, \mathfrak{R}_{2 \eta}(t_1,r_2,0,0)  
-\frac{t_2}{2\pi} \int_0^{2\pi} d r_2   \mathfrak{R}_{2 \eta}(t_1,r_2,0,0) 
\nonumber\\
&&+ \int_0^{t_3}  d r_3 \, \mathfrak{R}_{3 \eta}(t_1,t_2,r_3,0) 
-  \frac{t_3}{2\pi} \int_0^{2\pi} d r_3 \, \mathfrak{R}_{3 \eta}(t_1,t_2,r_3,0)  
\nonumber\\
&&+ \int_0^{t_4}  d r_4 \, \mathfrak{R}_{4 \eta}(t_1,t_2,t_3,r_4) 
- \frac{t_4}{2\pi} \int_0^{2\pi} d r_4  \, \mathfrak{R}_{4 \eta}(t_1,t_2,t_3,r_4) ,  
\end{eqnarray*}
and the bound follows from eq.~(\ref{eq:bound-R}).

\subsection{Proof of the lemma 2 }
\label{subsec:proof-lemma-4a}

We give a proof that the local currents, $j_\mu^{\diamond a}(x)$ and $j_\mu^\diamond(x)$,  
defined by eq.~(\ref{eq:measure-term-finite-volume})
satisfy all the properties required for the reconstruction theorem.  
Although the proof is  quite similar to that of theorem 5.3 in \cite{Luscher:1998du}, or 
that given in \cite{Kadoh:2007}, 
we give it here for completeness.  
\begin{enumerate}
\item {\it Smoothness.}  By construction, $j_\mu^{\diamond a}(x)$, $ j_\mu^\diamond(x)$ are 
defined for all admissible gauge fields in $\mathfrak{U}^{(2)}_Q \otimes \mathfrak{U}^{(1)}_{[0]}$. 
It depends smoothly on the link fields $U^{(2)}(x,\mu)$, 
$\tilde A_\mu^\prime(x)$ and $U_{[w]}(x,\mu)$ because 
$\hat P_-$ and $k_\mu$ are smooth functions of $U_s(x,\mu)$.  
Although $\tilde A_\mu^\prime(x)$ is not continuous 
when $\Lambda(x) = -1$ at some points $x$ because of  the cut in $\ln \Lambda(x)$,  its discontinuity is always 
in the pure-gauge form 
\begin{equation}
\text{disc.} \{ \tilde A_\mu^\prime(x) \} = - \partial_\mu \omega(x) ; \qquad \omega(0)  = 0,  
\end{equation}
where the gauge function $\omega(x)$ takes values that are integer multiples of $2\pi$.
Then, any smooth functionals of $\tilde A_\mu^\prime(x)$ are smooth with respect to the 
link field $U^{(1)}(x,\mu)$, if they are gauge-invariant under the gauge transformations 
$\tilde A^\prime_\mu(x) \rightarrow \tilde A^\prime_\mu(x) + \partial_\mu \omega(x)$ 
for arbitrary  periodic gauge functions $\omega(x)$ satisfying $\omega(0)=0$.
The currents $j_\mu^{\diamond a}(x)$ and $j_\mu^\diamond(x)$ are indeed gauge-invariant 
under such gauge transformations. 
Namely, 
taking the gauge covariance of $\hat P_-(x,y)$ and the gauge invariance of $k_\mu(x)$ into account,  
the change of  $\mathfrak{L}_\eta^\diamond$ under the gauge transformations 
\begin{eqnarray}
&& 
  \int_0^1 ds \, {\rm Tr} \big\{ \hat P_- \big[  [ \omega Y_-, \hat P_- ], \delta_\eta \hat P_-  \big] \big\}
+\int_0^1 ds \, {\rm Tr} \big\{ \hat P_+ \big[  [ \omega Y_+, \hat P_+ ], \delta_\eta \hat P_+  \big] \big\}
\nonumber\\
&&  
+ \int_0^1 ds \, \sum_{x \in \Gamma} \partial_\mu \omega(x) \, \delta_\eta \bar k_\mu(x)  \nonumber\\
&=& 
- \int_0^1 ds \, {\rm Tr} \big\{ \omega Y_- \, \delta_\eta \hat P_- \big\} 
- \int_0^1 ds \, {\rm Tr} \big\{ \omega Y_+ \, \delta_\eta \hat P_+ \big\} 
+ \int_0^1 ds \, \sum_{x \in \Gamma} \partial_\mu \omega(x) \, \delta_\eta k_\mu(x)  \nonumber\\
&=& \int_0^1 ds \sum_{x \in \Gamma} \omega(x) \, \delta_\eta\left\{ 
-{\rm tr}\{ Y_- \gamma_5 D \}(x,x) + {\rm tr}\{ Y_+ \gamma_5 D \}(x,x)  
- \partial^\ast_\mu k_\mu(x)  \right\}  = 0 , \nonumber\\
\end{eqnarray}
where the identity $\hat P_\pm \delta_\eta \hat P_\pm \hat P_\pm =0$ has been used.  


\item {\it Gauge invariance/covariance and symmetry properties.} The gauge invariance of 
$j_\mu^{\diamond a}(x)$ and $j_\mu^\diamond(x)$ under the U(1) gauge transformations
has been shown above.  
The transformation properties of $j_\mu^{\diamond a}(x)$, $ j_\mu^\diamond(x)$ 
under the SU(2) gauge transformations and the lattice symmetries are also evident from the transformation properties of 
$\hat P_-$, $k_\mu$ and ${\mathfrak{L}}_\eta \vert_{U=U^{(2)}\otimes U_{[w]}}$. 

\item {\it Integrability condition.}  From the definition of $\mathfrak{L}_\eta^\diamond$, 
eq.~(\ref{eq:measure-term-finite-volume}), one finds immediately that the second term does not contribute to the curvature 
$\delta_\eta \mathfrak{L}_\zeta^\diamond-\delta_\zeta \mathfrak{L}_\eta^\diamond
+\mathfrak{L}_{[\eta,\zeta]}^\diamond$ 
and the third term gives the curevature term at the gauge fields,  
$U^{(2)}(x,\mu)\otimes U_{[w]}(x,\mu)$.  
Taking the identity
${\rm Tr} \left\{ \delta_1 \hat P_\pm \delta_2 \hat P_\pm \delta_3\hat P_\pm \right\} =0$ 
into account, the curvature is evaluated as 
\begin{eqnarray}
\delta_\eta \mathfrak{L}_\zeta^\diamond-\delta_\zeta \mathfrak{L}_\eta^\diamond
+\mathfrak{L}_{[\eta,\zeta]}^\diamond
&=& 
 i \int_0^1 ds \, 
{\rm Tr} \left\{ 
  \hat P_- [ \delta_\eta \partial_s \hat P_-, \delta_\zeta \hat P_- ] 
-\hat P_- [ \delta_\zeta \partial_s \hat P_-, \delta_\eta \hat P_- ] 
\right\} \nonumber\\
&+& 
 i \int_0^1 ds \, 
{\rm Tr} \left\{ 
  \hat P_+ [ \delta_\eta \partial_s \hat P_+, \delta_\zeta \hat P_+ ] 
-\hat P_+ [ \delta_\zeta \partial_s \hat P_+, \delta_\eta \hat P_+ ] 
\right\} \nonumber\\
&+&
 \left. \left[
 i {\rm Tr} \left\{ \hat P_- [ \delta_\eta \hat P_-, \delta_\zeta \hat P_- ] \right\} 
 +
 i {\rm Tr} \left\{ \hat P_+ [ \delta_\eta \hat P_+, \delta_\zeta \hat P_+ ] \right\} 
\right]
 \right\vert_{U = U^{(2)}\otimes U_{[w]}} \nonumber\\
 &=&
 i \int_0^1 ds \,  \partial_s 
 \left[
 {\rm Tr} \left\{ 
  \hat P_- [ \delta_\eta  \hat P_-, \delta_\zeta \hat P_- ] 
 \right\} 
+
  {\rm Tr} \left\{ 
  \hat P_+ [ \delta_\eta  \hat P_+, \delta_\zeta \hat P_+ ] 
  \right\}
\right]
\nonumber\\
&+&
\left. \left[
 i {\rm Tr} \left\{ \hat P_- [ \delta_\eta \hat P_-, \delta_\zeta \hat P_- ] \right\} 
+ i {\rm Tr} \left\{ \hat P_+ [ \delta_\eta \hat P_+, \delta_\zeta \hat P_+ ] \right\} 
        \right]
 \right\vert_{U = U^{(2)}\otimes U_{[w]}}. \nonumber\\
\end{eqnarray}
After the integration in the first term,  the contributions from the lower end of the integration range 
cancels with the second term, 
because the variational parameters for the U(1) gauge field in this contribution
is restricted to $\eta_{\mu [w]}(x)$: 
\begin{equation}
\delta_{\eta^{(1)}} U_s(x,\mu)  \, U_s(x,\mu)^{-1}  \vert_{s=0} 
= [ s(\eta_\mu^{(1)}(x)- \eta_{\mu [w]}(x) ) + \eta_{\mu [w]}(x) ]_{s=0} = \eta_{\mu [w]}(x). 
\end{equation}
%

\item {\it Anomalous conservation law.}
If one sets $\eta_\mu^{(1)}(x) = - \partial_\mu \omega(x)$ 
(where $\omega(x)$ is any lattice function on $\Gamma$), 
the left-hand side of eq.~(\ref{eq:measure-term-finite-volume}) becomes
\begin{equation}
\sum_{x \in \Gamma} \omega(x) \, \partial_\mu^\ast  j_\mu^\diamond(x) . 
\end{equation}
On the other hand, using the identities
\begin{equation}
\delta_\eta \hat P_\pm = i s \left[ \omega Y_\pm , \hat P_\pm \right] , \qquad
\delta_\eta k_\mu(x) = 0, 
\end{equation}
the right-hand side is evaluated as 
\begin{eqnarray}
&& 
- \int_0^1 ds \, s \,  {\rm Tr} \big\{ \omega Y_+ \, \partial_s \hat P_+ \big\} 
- \int_0^1 ds \, s \,  {\rm Tr} \big\{ \omega Y_- \, \partial_s \hat P_- \big\} 
- \int_0^1 ds \, \sum_{x \in \Gamma} \partial_\mu \omega(x) \,  k_\mu(x)
\nonumber\\
&=& 
\sum_{x \in \Gamma} \omega(x) \, \left\{ 
{\rm tr}\{ Y_+ \gamma_5 D_L \}(x,x) - {\rm tr}\{ Y_- \gamma_5 D_L \}(x,x) \right\}
\nonumber\\
&+&
\int_0^1 ds \sum_{x \in \Gamma} \omega(x) \, \left\{ 
-{\rm tr}\{ Y_+ \gamma_5 D_L \}(x,x) +{\rm tr}\{ Y_+ \gamma_5 D_L \}(x,x) 
+ \partial_\mu^\ast k_\mu(x)  \right\} \nonumber\\
&=& 
\sum_{x \in \Gamma} \omega(x) \,
\left\{ -{\rm tr}\{ Y_+ \gamma_5(1- D_L) \}(x,x) +{\rm tr}\{ Y_+ \gamma_5(1- D_L) \}(x,x) \right\}. 
\end{eqnarray} 
Also, if one sets $\eta_\mu^{(2)}(x) = - \nabla_\mu \omega(x)$, 
the left-hand side of eq.~(\ref{eq:measure-term-finite-volume}) becomes
\begin{equation}
\sum_{x \in \Gamma} \omega^a(x) \, \{ \nabla_\mu^\ast  j_\mu^\diamond \}^a(x) . 
\end{equation}
On the other hand, using the identities
\begin{equation}
\delta_\eta \hat P_- = i  \left[ \omega  , \hat P_- \right] , \qquad
\delta_\eta k_\mu(x) = 0, 
\end{equation}
the right-hand side is evaluated as 
\begin{eqnarray}
&& 
- \int_0^1 ds \, \partial_s \,  {\rm Tr} \big\{ \omega \hat P_- \big\} 
+{\mathfrak{L}}_\eta \vert_{U=U^{(2)} \otimes U_{[w]};\eta_\mu^{(2)} = - \nabla_\mu \omega}
\nonumber\\
&=& 
- \sum_{x \in \Gamma} \omega^a(x) \,  {\rm tr}\{ T^a \gamma_5 D_L \}(x,x) \nonumber\\
&&+
\sum_{x \in \Gamma} \omega^a(x) \, \left\{
 {\rm tr}\{ T^a \gamma_5 D_L \}(x,x) + \{ \nabla_\mu j_\mu^\diamond \}^a(x) 
 \right\}\vert_{U=U^{(2)} \otimes U_{[w]}}. 
\end{eqnarray} 
The last term vanishes identically if the anomalous conservation laws hold 
for the measure term ${\mathfrak{L}}_\eta \vert_{U=U^{(2)} \otimes U_{[w]}}$ 
at the gauge fields $U^{(2)}(x,\mu) \otimes U_{[w]}(x,\mu)$. This follows from its definition
eq.~(\ref{eq:measure-term-Wilson-lines-su2}) by noting 
\begin{equation}
\mathfrak{C}_{\nu \eta}(t) \vert_{\eta_\mu^{(2)} = - \nabla_\mu \omega}=
-\partial_{t_\nu} \,  {\rm Tr} \big\{ \omega \hat P_- \big\}(t), \qquad
\delta_\eta  \phi_{[w]}  \vert_{\eta_\mu^{(2)} = - \nabla_\mu \omega}=0, 
\end{equation}
and the fact that the SU(2) gauge anomaly ${\rm tr}\{ T^a \gamma_5(1- D_L) \}(x,x) $ vanishes 
identically when the U(1) gauge field is trivial $(t=0)$ due to the pseudo reality of SU(2). 

\end{enumerate}

\subsection{Locality properties of the measure term currents}
\label{sec:locality-of-currents}

Finally, we examine the locality property of the measure term 
currents, $j_\mu^{\diamond a}(x)$ and $j_\mu^\diamond(x)$.  
We follow  the procedure to decompose 
the measure term eq.~(\ref{eq:measure-term-finite-volume}) into the part definable in infinite volume 
and the part of the finite volume corrections. 
Namely, the measure term eq.~(\ref{eq:measure-term-finite-volume}) may be decomposed as follows:
\begin{equation}
{\mathfrak{L}}_\eta^\diamond = {\mathfrak{K}}_\eta^\diamond  + {\mathfrak{S}}_\eta^\diamond , 
\end{equation}
where 
\begin{eqnarray}
\label{eq:measure-term-finite-volume-infinite-part}
\mathfrak{K}_\eta^\diamond &=& 
i \int_0^1 ds \, 
{\rm Tr} \left\{ Q_\Gamma \hat P_- [ \partial_s \hat P_-,  \delta_\eta \hat P_- ] \right\} 
+
i \int_0^1 ds \, 
{\rm Tr} \left\{ Q_\Gamma \hat P_+ [ \partial_s \hat P_+,  \delta_\eta \hat P_+ ] \right\} 
\nonumber\\
&& \qquad \qquad 
+
\delta_\eta \, 
\int_0^1 ds \, \sum_{x \in \Gamma} 
 \left\{   \tilde A_\mu^\prime(x)  \,   \bar k_\mu(x) \right\} ,  \\
\label{eq:measure-term-finite-volume-finite-part}
\mathfrak{S}_\eta^\diamond &=& \,  
\int_0^1 ds  \, \, 
\mathfrak{R}_{\zeta \eta}\vert_{\zeta_\mu = \tilde A_\mu^\prime} \nonumber\\
&&  \qquad \qquad
+\delta_\eta \, 
\int_0^1 ds \, \sum_{x \in \Gamma} 
 \left\{   \tilde A_\mu^\prime(x)  \,   \Delta k_\mu(x) \right\}   
+{\mathfrak{L}}_\eta \vert_{U=U^{(2)} \otimes U_{[w]}} . 
\end{eqnarray}
From the bounds 
eqs.~(\ref{eq:bound-delta-kmu}), (\ref{eq:bound-R}) 
and eq.~(\ref{eq:measure-term-Wilson-lines-bound2}) and 
$\| A_\mu^T(x) \| \le \kappa_6 L^4 \, (\kappa_6 >0)$ \cite{Luscher:1998du}, one can infer 
\begin{equation}
\label{eq:bound-S-diamond}
 \left| \mathfrak{S}^\diamond_{\eta} \right|  \le 
\kappa_3 L^{\nu_3} {\rm e}^{-L/\varrho} \, \| \eta \|_\infty 
\end{equation}
for some constants $\kappa_3 >0$, $\nu_3 \ge 0$.

As to ${\mathfrak{K}}_\eta^\diamond$ defined by  
eq.~(\ref{eq:measure-term-finite-volume-infinite-part}), 
if one introduces the truncated fields
\begin{equation}
\eta^n_\mu(x) = \left\{ \begin{array}{cl} \eta_\mu(x) & \text{if $x-Ln \in \Gamma$}, \\
                                                                     0 & \text{otherwise, } \end{array} \right. 
\end{equation}
for any integer vector $n$, 
it may be rewritten into 
\begin{eqnarray}
\label{eq:measure-term-finite-volume-infinite-part-2}
\mathfrak{K}_\eta^\diamond &=&
 i \int_0^1 ds \, {\rm Tr} \left\{ P_- [ \partial_s P_-, \delta_{\eta^0} P_- ] \right\} 
+ i \int_0^1 ds \, {\rm Tr} \left\{ P_+ [ \partial_s P_+ \delta_{\eta^0} P_+] \right\} 
\nonumber\\
&& \qquad \quad 
+
\int_0^1 ds \, \sum_{x \in \mathbb{Z}^4} \left\{ 
( \eta_\mu^{(1) 0}(x)-\eta_{\mu[w]}^0(x) )\, \bar k_\mu(x) 
+  \tilde A_\mu^\prime(x)  \,   \delta_{\eta^0} \, \bar k_\mu(x) \right\}. 
\end{eqnarray}
One can see from this expression
that $\mathfrak{K}_\eta^\diamond$ is defined in infinite volume for  
the variational parameter with a compact support $\eta_\mu^0(x)$. 
Then the following lemma holds ture:

\vspace{1.5em}
\noindent{\bf Lemma 4} \ \ {\sl ${\mathfrak{K}}_\eta^\diamond$ is in the form
\begin{equation}
\label{eq:K-to-L-star}
{\mathfrak{K}}_\eta^\diamond = \mathfrak{L}^\star_{\eta^0}, 
\end{equation}
where $\mathfrak{L}^\star_{\eta}$ is the linear functional defined in infinite volume 
for any variation parameter $\eta_\mu(x)$ with a compact support given by
\begin{eqnarray}
\label{eq:measure-term-infinite-volume-star}
\mathfrak{L}_{\eta}^\star &=&
     i \int_0^1 ds \,\Big[ {\rm Tr} \left\{ P_- [ \partial_s P_-, \delta_{\eta} P_- ] \right\} 
                                  +  {\rm Tr} \left\{ P_+ [ \partial_s P_+, \delta_{\eta} P_+ ] \right\} 
   \Big]_{U_s=U^{(2)} \otimes \, {\rm e}^{i  s \tilde A_\mu} } 
\nonumber\\
&&  \qquad \, 
+\int_0^1 ds \, \Big[
\sum_{x \in \mathbb{Z}^4} \Big\{ 
\eta_\mu^{(1)}(x) \, \bar k_\mu(x) 
+  \tilde A_\mu(x)  \,   \delta_{\eta} \, \bar k_\mu(x) \Big\}
\Big]_{U_s=U^{(2)} \otimes \, {\rm e}^{i  s \tilde A_\mu} } 
\nonumber\\
&\equiv& \sum_{x \in \mathbb{Z}^4}  
\{ \eta^a_\mu(x) j^{a  \star}_\mu(x)+ \eta_\mu(x) j_\mu^\star(x) \}.  
\end{eqnarray}
$\tilde A_\mu(x)$ here 
is the vector potential (in infinite volume) which represents the 
U(1) link field in the topological sector $\mathfrak{U}^{(1)}[0]$ 
(periodic in infinite volume),  with the following properties, 
\begin{eqnarray}
\label{eq:prop-tilde-A}
&& U^{(1)}(x,\mu) = {\rm e}^{i \tilde A_\mu(x) },  
\quad   | \tilde A_\mu(x) |  \le \pi( 1 + 4 \| x \| ) , \nonumber\\
&& F_{\mu\nu}(x) = \partial_\mu \tilde A_\nu(x)- \partial_\nu \tilde A_\mu(x)
\end{eqnarray}
and any other field with these properties is equal to $\tilde A_\mu(x)+\partial_\mu \omega(x)$, 
where the gauge function $\omega(x)$ takes values that are integer multiples of $2\pi$.
}

\vspace{2em}
\noindent
The proof of this lemma has been given for the U(1) case in our previous work \cite{Kadoh:2007}
and it applies to the SU(2) $\times$ U(1) case here simply by regarding the SU(2) link field as a background. So we omit it here. 

The currents $j^{a  \star}_\mu(x)$ and  $j_{\mu}^\star(x)$ are quite similar in construction 
to $j_\mu^\star(x)$ defined in  \cite{Luscher:1998du} for the U(1) case. 
In particular, 
they are invariant under the gauge transformations
$\tilde A_\mu(x) \rightarrow \tilde A_\mu(x)+\partial_\mu \omega(x)$ for arbitrary  gauge functions 
$\omega(x)$ that are  polynomially bounded at infinity.  Then, 
the locality property of $j^{a  \star}_\mu(x)$ and  $j_{\mu}^\star(x)$ 
with respect to the U(1) link field can be established by the same 
argument as that given in  \cite{Luscher:1998du}.  The locality property with respect to 
the SU(2) link field follows from the locality property of the kernels of projection operators $P_\pm(x,y)$ and the current $\bar k_\mu(x)$.

\section{Measure term in infinite volume} 
\label{sec:measure-term-in-infinite-volume} 

We note that $\mathfrak{L}_{\eta}^\star$ defined by eq.~(\ref{eq:measure-term-infinite-volume-star})
provides the measure term of the lattice Glashow-Weinberg-Salam model in infinite volume. 
This non-perturbative result should be compared with the construction of the measure term 
in the weak coupling expansion \cite{Luscher:2000zd}.  One can see through  the weak coupling expansion of 
eq.~(\ref{eq:measure-term-infinite-volume-star}) that 
the result of \cite{Luscher:2000zd} is automatically reproduced for the case of the SU(2)$\times$U(1) chiral gauge theory. 
Thus it provides a gauge-invariant lattice regularization of the Glashow-Weinberg-Salam model 
to all orders of  perturbation theory.

\section{Discussion}
\label{sec:discussion}

In this paper, we have given a gauge-invariant and non-perturbative construction of 
the Glashow-Weinberg-Salam model on the lattice, based on the lattice Dirac operator
satisfying the Ginsparg-Wilson relation. 
We have shown that it is indeed possible to construct 
the fermion measure of quarks and leptons
which depends smoothly on the SU(2)$\times$U(1) gauge fields  and 
fulfills the fundamental requirements such as locality,  gauge-invariance and lattice symmetries
in all SU(2) topological sectors with vanishing U(1) magnetic flux. 
Then this construction would be usable for the studies of non-perturbative aspects of the 
Glashow-Weinberg-Salam model, such as the baryon number non-conservation. 
However,  it is still desirable to extend our result in this paper to the topological sectors 
with non-vanishing U(1) magnetic fluxes. 

The measure term for 
the SU(2)$\times$U(1) chiral gauge theory of the  Glashow-Weinberg-Salam model 
may be constructed by solving the local cohomology problem formulated in 4+2 dimensions 
for generic non-abelian gauge theories \cite{Luscher:1999un,Adams:2000yi,Alvarez-Gaume:1983cs}.  The problem has been solved only in the infinite volume limit so far \cite{Kikukawa:2000kd}. 
The measure term obtained in this paper
provides an explicit solution 
to the 4+2 dimensional local cohomology problem in the finite volume 
for the topological sectors with vanishing U(1) magnetic fluxes. 

As for the formulation in the infinite volume, one may adopt the non-compact formulation for the U(1) gauge theory, as discussed by Neuberger in \cite{Neuberger:2000wq}. Even for this case, 
the expression of the measure term given by eq.~(\ref{eq:measure-term-infinite-volume-star}) holds true, 
if the vector potential there is identified as the dynamical field variables in the non-compact U(1) formulation. 

Towards a numerical  application of the SU(2)$\times$U(1) chiral lattice gauge theory 
of the Glashow-Weinberg-Salam model, 
the next step is the practical implementation of the formula 
of the chiral bases, eqs.~(\ref{eq:chiral-basis-choice-v})-(\ref{eq:evolution-operators}): 
a computation of $W$ and 
the implementation of the operator $Q_{t \pm}$. 
This question has been addressed partly  for the U(1) case in our previous 
works \cite{Kadoh:2004uu, Kikukawa:2001mw, Aoyama:1999hg}. 
We will disscuss this question in detail elsewhere.

\bigskip

\acknowledgments

Y.K. would like to thank Ting-Wai Chiu for his kind hospitality at 2005 Taipei
Summer Institute on Strings, Particles and Fields and 
D.~Adams, K.~Fujikawa,  H.~Suzuki for discussions. 
Y.K.  is also grateful to Y.~Kuramashi for his nice organization of ILFTN workshop on 
"Perspectives in Lattice QCD" (Nara, 2005). Some of the proofs given in this paper 
were completed during this workshop. 
Y.K. is supported in part by Grant-in-Aid for Scientific Research No.~17540249.


\begin{thebibliography}{999}

\bibitem{'tHooft:1976up}
  G.~'t Hooft,
  Phys.\ Rev.\ Lett.\  {\bf 37}, 8 (1976).
  
\bibitem{'tHooft:1976fv}
  G.~'t Hooft,
  Phys.\ Rev.\ D {\bf 14}, 3432 (1976)
  [Erratum-ibid.\ D {\bf 18}, 2199 (1978)].
   
\bibitem{Raby:1979my}
  S.~Raby, S.~Dimopoulos and L.~Susskind,
  Nucl.\ Phys.\ B {\bf 169}, 373 (1980).
  
%

\bibitem{Dimopoulos:1980hn}
  S.~Dimopoulos, S.~Raby and L.~Susskind,
  Nucl.\ Phys.\ B {\bf 173}, 208 (1980).

\bibitem{'tHooft:1979bh}
  G.~'t Hooft,
PRINT-80-0083 (UTRECHT)
{\it Lecture given at Cargese Summer Inst., Cargese, France, Aug 26 - Sep 8, 1979}


\bibitem{Karsten:1980wd}
  L.~H.~Karsten and J.~Smit,
  Nucl.\ Phys.\  B {\bf 183}, 103 (1981).

\bibitem{Nielsen:1980rz}
  H.~B.~Nielsen and M.~Ninomiya,
  Nucl.\ Phys.\  B {\bf 185}, 20 (1981)
  [Erratum-ibid.\  B {\bf 195}, 541 (1982)].

\bibitem{Nielsen:1981xu}
  H.~B.~Nielsen and M.~Ninomiya,
  Nucl.\ Phys.\  B {\bf 193}, 173 (1981).

\bibitem{Friedan:1982nk}
  D.~Friedan,
  Commun.\ Math.\ Phys.\  {\bf 85}, 481 (1982).
  


\bibitem{Ginsparg:1981bj}
P.~H.~Ginsparg and K.~G.~Wilson,
Phys.\ Rev.\ D {\bf 25}, 2649 (1982).


\bibitem{Neuberger:1997fp}
H.~Neuberger,
Phys.\ Lett.\ B {\bf 417}, 141 (1998)
[arXiv:hep-lat/9707022].

\bibitem{Hasenfratz:1998ri}
P.~Hasenfratz, V.~Laliena and F.~Niedermayer,
Phys.\ Lett.\ B {\bf 427}, 125 (1998)
[arXiv:hep-lat/9801021].

\bibitem{Neuberger:1998wv}
H.~Neuberger,
Phys.\ Lett.\ B {\bf 427}, 353 (1998)
[arXiv:hep-lat/9801031].

\bibitem{Hasenfratz:1998jp}
P.~Hasenfratz,
Nucl.\ Phys.\ B {\bf 525}, 401 (1998)
[arXiv:hep-lat/9802007].


\bibitem{Hernandez:1998et}
P.~Hernandez, K.~Jansen and M.~L\"uscher,
Nucl.\ Phys.\ B {\bf 552}, 363 (1999)
[arXiv:hep-lat/9808010].

%


\bibitem{Luscher:1998pq}
M.~L\"uscher,
Phys.\ Lett.\ B {\bf 428}, 342 (1998)
[arXiv:hep-lat/9802011].


\bibitem{Luscher:1998kn}
M.~L\"uscher,
Nucl.\ Phys.\ B {\bf 538}, 515 (1999)
[arXiv:hep-lat/9808021].

\bibitem{Luscher:1998du}
M.~L\"uscher,
Nucl.\ Phys.\ B {\bf 549}, 295 (1999)
[arXiv:hep-lat/9811032].

\bibitem{Luscher:1999un}
M.~L\"uscher,
Nucl.\ Phys.\ B {\bf 568}, 162 (2000)
[arXiv:hep-lat/9904009].

\bibitem{Luscher:1999mt}
M.~L\"uscher,
Nucl.\ Phys.\ Proc.\ Suppl.\  {\bf 83}, 34 (2000)
[arXiv:hep-lat/9909150].

\bibitem{Luscher:2000hn}
M.~L\"uscher,
arXiv:hep-th/0102028.

\bibitem{Suzuki:1999qw}
H.~Suzuki,
Prog.\ Theor.\ Phys.\  {\bf 101}, 1147 (1999)
[arXiv:hep-lat/9901012].

\bibitem{Neuberger:2000wq}
H.~Neuberger,
Phys.\ Rev.\ D {\bf 63}, 014503 (2001)
[arXiv:hep-lat/0002032].


\bibitem{Adams:2000yi}
D.~H.~Adams,
Nucl. Phys. B {\bf 589}, 633 (2000)
[arXiv:hep-lat/0004015].

\bibitem{Suzuki:2000ii}
H.~Suzuki,
Nucl.\ Phys.\ B {\bf 585}, 471 (2000)
[arXiv:hep-lat/0002009].

\bibitem{Igarashi:2000zi}
H.~Igarashi, K.~Okuyama and H.~Suzuki,
arXiv:hep-lat/0012018.

\bibitem{Luscher:2000zd}
M.~L\"uscher,
JHEP {\bf 0006}, 028 (2000)
[arXiv:hep-lat/0006014].

\bibitem{Kikukawa:2000kd}
Y.~Kikukawa and Y.~Nakayama,
Nucl.\ Phys.\ B {\bf 597}, 519 (2001)
[arXiv:hep-lat/0005015].

\bibitem{Kikukawa:2001jm}
  Y.~Kikukawa, Y.~Nakayama and H.~Suzuki,
  Nucl.\ Phys.\ Proc.\ Suppl.\  {\bf 106}, 763 (2002)
  [arXiv:hep-lat/0111036].
  
\bibitem{Glashow:1961tr}
  S.~L.~Glashow,
  Nucl.\ Phys.\  {\bf 22}, 579 (1961).
  
\bibitem{Weinberg:1967tq}
  S.~Weinberg,
  Phys.\ Rev.\ Lett.\  {\bf 19}, 1264 (1967).
  
\bibitem{Salam:1968rm}
  A.~Salam,
{\it  Originally printed in *Svartholm: Elementary Particle Theory, Proceedings Of The Nobel Symposium Held 1968 At Lerum, Sweden*, Stockholm
1968, 367-377}

\bibitem{Kadoh:2003ii}
 D.~Kadoh, Y.~Kikukawa and Y.~Nakayama,
JHEP {\bf 0412}, 006 (2004)
[arXiv:hep-lat/0309022].
  
\bibitem{Kadoh:2004uu}
  D.~Kadoh and Y.~Kikukawa,
  JHEP {\bf 0501}, 024 (2005)
  [arXiv:hep-lat/0401025].

\bibitem{Kadoh:2007}
 D.~Kadoh and Y.~Kikukawa,
``A simple construction of fermion measure term in U(1) chiral lattice gauge theories
with exact gauge invariance'',  arXiv:0709.3656.

%

\bibitem{Narayanan:wx}
R.~Narayanan and H.~Neuberger,
Phys.\ Lett.\ B {\bf 302}, 62 (1993)
[arXiv:hep-lat/9212019].
%

\bibitem{Narayanan:sk}
R.~Narayanan and H.~Neuberger,
Nucl.\ Phys.\ B {\bf 412}, 574 (1994)
[arXiv:hep-lat/9307006].

\bibitem{Narayanan:ss}
R.~Narayanan and H.~Neuberger,
Phys.\ Rev.\ Lett.\  {\bf 71}, 3251 (1993)
[arXiv:hep-lat/9308011].

\bibitem{Narayanan:1994gw}
R.~Narayanan and H.~Neuberger,
Nucl.\ Phys.\ B {\bf 443}, 305 (1995)
[arXiv:hep-th/9411108].

\bibitem{Narayanan:1993gq}
R.~Narayanan,
Nucl.\ Phys.\ Proc.\ Suppl.\  {\bf 34}, 95 (1994)
[arXiv:hep-lat/9311014].

\bibitem{Neuberger:1999ry}
H.~Neuberger,
Nucl.\ Phys.\ Proc.\ Suppl.\  {\bf 83}, 67 (2000)
[arXiv:hep-lat/9909042].

\bibitem{Narayanan:1996cu}
R.~Narayanan and H.~Neuberger,
Nucl.\ Phys.\ B {\bf 477}, 521 (1996)
[arXiv:hep-th/9603204].

\bibitem{Huet:1996pw}
P.~Y.~Huet, R.~Narayanan and H.~Neuberger,
Phys.\ Lett.\ B {\bf 380}, 291 (1996)
[arXiv:hep-th/9602176].

\bibitem{Narayanan:1997by}
R.~Narayanan and J.~Nishimura,
Nucl.\ Phys.\ B {\bf 508}, 371 (1997)
[arXiv:hep-th/9703109].


\bibitem{Kikukawa:1997qh}
Y.~Kikukawa and H.~Neuberger,
Nucl.\ Phys.\ B {\bf 513}, 735 (1998)
[arXiv:hep-lat/9707016].

\bibitem{Neuberger:1998xn}
  H.~Neuberger,
  Phys.\ Rev.\ D {\bf 59}, 085006 (1999)
  [arXiv:hep-lat/9802033].



\bibitem{Kaplan:1992bt}
D.~B.~Kaplan,
Phys.\ Lett.\ B {\bf 288}, 342 (1992)
[arXiv:hep-lat/9206013].

\bibitem{Shamir:1993zy}
Y.~Shamir,
Nucl.\ Phys.\ B {\bf 406}, 90 (1993)
[arXiv:hep-lat/9303005].

\bibitem{Furman:ky}
V.~Furman and Y.~Shamir,
Nucl.\ Phys.\ B {\bf 439}, 54 (1995)
[arXiv:hep-lat/9405004].

\bibitem{Blum:1996jf}
T.~Blum and A.~Soni,
Phys.\ Rev.\ D {\bf 56}, 174 (1997)
[arXiv:hep-lat/9611030].

\bibitem{Blum:1997mz}
T.~Blum and A.~Soni,
Phys.\ Rev.\ Lett.\  {\bf 79}, 3595 (1997)
[arXiv:hep-lat/9706023].

\bibitem{Vranas:1997da}
P.~M.~Vranas,
Phys.\ Rev.\ D {\bf 57}, 1415 (1998)
[arXiv:hep-lat/9705023].

\bibitem{Neuberger:1997bg}
H.~Neuberger,
Phys.\ Rev.\ D {\bf 57}, 5417 (1998)
[arXiv:hep-lat/9710089].

\bibitem{Kikukawa:1999sy}
Y.~Kikukawa and T.~Noguchi,
arXiv:hep-lat/9902022.


%
\bibitem{Narayanan:1996kz}
R.~Narayanan and H.~Neuberger,
Phys.\ Lett.\ B {\bf 393}, 360 (1997)
[Phys.\ Lett.\ B {\bf 402}, 320 (1997)]
[arXiv:hep-lat/9609031].

\bibitem{Kikukawa:1997md}
Y.~Kikukawa, R.~Narayanan and H.~Neuberger,
Phys.\ Lett.\ B {\bf 399}, 105 (1997)
[arXiv:hep-th/9701007].

\bibitem{Kikukawa:1997dv}
Y.~Kikukawa, R.~Narayanan and H.~Neuberger,
Phys.\ Rev.\ D {\bf 57}, 1233 (1998)
[arXiv:hep-lat/9705006].



\bibitem{Kikukawa:2001mw}
Y.~Kikukawa,
Phys.\ Rev.\ D {\bf 65}, 074504 (2002)
[arXiv:hep-lat/0105032].

\bibitem{Aoyama:1999hg}
T.~Aoyama and Y.~Kikukawa,
arXiv:hep-lat/9905003.


\bibitem{Creutz:1996xc}
  M.~Creutz, M.~Tytgat, C.~Rebbi and S.~S.~Xue,
  Phys.\ Lett.\  B {\bf 402}, 341 (1997)
  [arXiv:hep-lat/9612017].

\bibitem{Eichten:1985ft}
  E.~Eichten and J.~Preskill,
  Nucl.\ Phys.\  B {\bf 268}, 179 (1986).
   
\bibitem{Bhattacharya:2006dc}
  T.~Bhattacharya, M.~R.~Martin and E.~Poppitz,
  Phys.\ Rev.\  D {\bf 74}, 085028 (2006)
  [arXiv:hep-lat/0605003].

\bibitem{Giedt:2007qg}
  J.~Giedt and E.~Poppitz,
  arXiv:hep-lat/0701004.

\bibitem{Poppitz:2007tu}
  E.~Poppitz and Y.~Shang,
  arXiv:0706.1043 [hep-th].
 
\bibitem{Gerhold:2007yb}
  P.~Gerhold and K.~Jansen,
  JHEP {\bf 0709}, 041 (2007)
  [arXiv:0705.2539 [hep-lat]].

\bibitem{Gerhold:2007gx}
  P.~Gerhold and K.~Jansen,
  arXiv:0707.3849 [hep-lat].


\bibitem{Kikukawa:2007im}
  Y.~Kikukawa and H.~Suzuki,
  arXiv:0708.1989 [hep-lat].
  


 
\bibitem{Suzuki:2000ku}
  H.~Suzuki,
  JHEP {\bf 0010}, 039 (2000)
  [arXiv:hep-lat/0009036].
  
\bibitem{Fujikawa:2002vj}
  K.~Fujikawa, M.~Ishibashi and H.~Suzuki,
  JHEP {\bf 0204}, 046 (2002)
  [arXiv:hep-lat/0203016].
  
\bibitem{Fujikawa:2002up}
  K.~Fujikawa and H.~Suzuki,
  Phys.\ Rev.\ D {\bf 67}, 034506 (2003)
  [arXiv:hep-lat/0210013].

\bibitem{Giusti:2002rx}
  L.~Giusti,
  Nucl.\ Phys.\ Proc.\ Suppl.\  {\bf 119}, 149 (2003)
  [arXiv:hep-lat/0211009].

\bibitem{Kobayashi:1973fv}
  M.~Kobayashi and T.~Maskawa,
  Prog.\ Theor.\ Phys.\  {\bf 49}, 652 (1973).

\bibitem{Fujiwara:1999fj}
T.~Fujiwara, H.~Suzuki and K.~Wu,
Phys.\ Lett.\ B {\bf 463}, 63 (1999)
[arXiv:hep-lat/9906016].




\bibitem{Adams:2002ms}
  D.~H.~Adams,
  Nucl.\ Phys.\  B {\bf 640}, 435 (2002)
  [arXiv:hep-lat/0203014].
  
 
\bibitem{Neuberger:1998rn}
  H.~Neuberger,
  Phys.\ Lett.\  B {\bf 437}, 117 (1998)
  [arXiv:hep-lat/9805027].
  
\bibitem{Bar:2002sa}
  O.~Bar,
  Nucl.\ Phys.\  B {\bf 650}, 522 (2003)
  [arXiv:hep-lat/0209098].

\bibitem{Bar:2000qa}
  O.~Bar and I.~Campos,
  Nucl.\ Phys.\  B {\bf 581}, 499 (2000)
  [arXiv:hep-lat/0001025].

\bibitem{Kikukawa:1998pd}
Y.~Kikukawa and A.~Yamada,
Phys.\ Lett.\ B {\bf 448}, 265 (1999)
[arXiv:hep-lat/9806013].

\bibitem{Fujikawa:1998if}
K.~Fujikawa,
Nucl.\ Phys.\ B {\bf 546}, 480 (1999)
[arXiv:hep-th/9811235].

\bibitem{Adams:1998eg}
D.~H.~Adams,
Annals Phys.\  {\bf 296}, 131 (2002)
[arXiv:hep-lat/9812003].

\bibitem{Suzuki:1998yz}
H.~Suzuki,
Prog.\ Theor.\ Phys.\  {\bf 102}, 141 (1999)
[arXiv:hep-th/9812019].

\bibitem{Chiu:1998xf}
T.~W.~Chiu,
Phys.\ Lett.\ B {\bf 445}, 371 (1999)
[arXiv:hep-lat/9809013].



\bibitem{Fujiwara:1999fi}
T.~Fujiwara, H.~Suzuki and K.~Wu,
Nucl.\ Phys.\ B {\bf 569}, 643 (2000)
[arXiv:hep-lat/9906015].

\bibitem{Igarashi:2002zz}
H.~Igarashi, K.~Okuyama and H.~Suzuki,
Nucl.\ Phys.\ B {\bf 644}, 383 (2002)
[arXiv:hep-lat/0206003].


\bibitem{Alvarez-Gaume:1983cs}
L.~Alvarez-Gaume and P.~H.~Ginsparg,
Nucl. Phys. B {\bf 243}, 449 (1984).
%



%
%

%
%

%

%

\end{thebibliography}
\end{document}